\documentclass[prd,amsmath,amssymb,floatfix,superscriptaddress,notitlepage,nofootinbib,preprintnumbers]{revtex4-1}
\usepackage{amsfonts,amssymb,amsmath,graphicx,color,bm,enumitem}
\definecolor{ultramarine}{rgb}{0.07, 0.04, 0.56}
\definecolor{cadmiumgreen}{rgb}{0.0, 0.42, 0.24}
\definecolor{indigo(dye)}{rgb}{0.0, 0.25, 0.42}
\usepackage[linktocpage=true]{hyperref}
\hypersetup{
colorlinks=true,
citecolor=blue,
linkcolor=red,
urlcolor=indigo(dye),
}

\newcommand{\be}{\begin{equation}}  
\newcommand{\ee}{\end{equation}}
\newcommand{\bem}{\begin{pmatrix}}
\newcommand{\eem}{\end{pmatrix}}

\begin{document}

\title{Stealth spontaneous spinorization of relativistic stars}

\author{Masato Minamitsuji}
\affiliation{Center for Astrophysics and Gravitation (CENTRA), Instituto Superior T\'ecnico, University of Lisbon,
Lisbon 1049-001, Portugal.}

\begin{abstract}
We investigate the behavior of the Dirac spinor fields in general relativistic high density stellar backgrounds and the possibility of spontaneous spinorization which is analogous to spontaneous scalarization. We consider the model with  the modified kinetic term of the Dirac field by the insertion of the fifth gamma matrix ${\hat \gamma}^5$ and the conformal coupling of the Dirac spinor field to the matter sector, which would lead to the tachyonic growth of the Dirac spinor field in the high density compact stellar backgrounds. In order to obtain the static and spherically symmetric solutions, we have to consider the two Dirac fields at the same time. We show that in the constant density stellar backgrounds our model gives rise to the nontrivial solutions of the Dirac spinor fields with any number of nodes, where one mode has one more node than the other. We also show that at the leading order all the components of the effective energy-momentum tensor of the Dirac spinor fields, after the summation over the two  fields, vanish for any separable time-dependent ansatz of the Dirac spinor fields in any static and spherically symmetric spacetime backgrounds, indicating that spontaneous spinorization takes place as a stealth process in the static and spherically symmetric spacetime with any number of nodes, which would be quenched by nonlinear effects and leave no observable effects.
\end{abstract}
\pacs{04.50.-h, 04.50.Kd, 98.80.-k}
\keywords{Higher-dimensional Gravity, Modified Theories of Gravity, Cosmology}
\maketitle

\section{Introduction}
\label{sec1}

Einstein's general relativity
has been tested 
by various  experiments and observations
on the scales
from the Solar System to binary pulsars~\cite{Will:2014kxa},
with the dawn of gravitational wave observations,
and the frontier for testing genetal relativity
will expand to strong gravity regimes \cite{Berti:2015itd,Berti:2018vdi,Berti:2018cxi}.
A potentially interesting phenomenon
which could be tested in strong gravity regimes
is spontaneous scalarization
which allows for large deviations from general relativity 
via the tachyonic growth of the scalar field 
inside relativistic stars~\cite{Damour:1993hw}.
In scalar-tensor systems with the conformal coupling to matter,
a stellar solution in general relativity with the vanishing scalar field $\phi=0$
suffers from a tachyonic instability, 
leading to a relativistic star solution
with a nontrivial profile of the scalar field 
~\cite{Damour:1996ke,Harada:1997mr,Harada:1998ge,Novak:1998rk,Palenzuela:2013hsa,Sampson:2014qqa,Pani:2014jra,Silva:2014fca}.

The idea of spontaneous scalarization may be extended
to other field species such as vector and spinor fields.
The possibility of spontaneous vectorization
triggered by the tachyonic instability of the vector field
has been studied recently
in Refs. \cite{Ramazanoglu:2017xbl,Ramazanoglu:2019gbz,Annulli:2019fzq,Ramazanoglu:2019jrr,Kase:2020yhw,Minamitsuji:2020pak}.
Refs.~\cite{Annulli:2019fzq,Kase:2020yhw}
have studied the model 
with nonminimal couplings of the vector field to the spacetime curvature,
and  constructed static and spherically symmetric relativistic star solutions 
with nontrivial vector field profiles, 
which would be formed via the selected choice of the initial conditions
rather via the instability of the general relativistic solutions.
On the other hand, 
the analyses 
from the Jordan frame viewpoint
have been presented in Refs. \cite{Ramazanoglu:2017xbl,Minamitsuji:2020pak},
and showed that spontaneous vectorization could happen
as in the same manner as spontaneous scalarization
via the vector conformal coupling to matter.

In this paper,
we will explore the similar phenomena in the fermionic sectors,
namely, {\it spontaneous spinorization}.
This possibility was discussed \cite{Ramazanoglu:2018hwk} at the first time.
For clarification,
we start with the theory of the standard massive Dirac spinor field  $\psi$ 
in the Minkowski spacetime given by  
\begin{eqnarray}
\label{dirac}
S_{(\psi)}
=
\int d^4X
\left[
\frac{1}{2}
\left(
{\bar \psi}
{\hat\gamma}^a
\partial_a
\psi
-
\left(
\partial_a
{\bar \psi}
\right)
{\hat\gamma}^a 
\psi  
\right)
-m 
{\bar \psi}
\psi
\right],
\end{eqnarray}
where 
the indices $a, b, c, \cdots=0,1,2,3$ run
the four-dimensional Minkowski spacetime with the metric $\eta_{ab}$,
\begin{eqnarray}
\label{reference}
d{\hat s}^2
:=
\eta_{ab}dX^a dX^b
=
-d(X^0)^2
+d(X^1)^2
+d(X^2)^2
+d(X^3)^2,
\end{eqnarray}
$\partial_a:=\partial/\partial X^a$
is the partial derivative
with respect to the coordinate $X^a$,
$m$ is the constant mass parameter,
the Dirac gamma matrices in the Minkowski spacetime ${\hat \gamma}^a$
satisfy the anticommutation relations
\begin{eqnarray}
\label{dmatrix}
\{
{\hat\gamma}^a,
{\hat \gamma}^b
\}
=2\eta^{ab}
I_{4\times 4},
\end{eqnarray}
with $I_{4\times 4}$ being the $4\times 4$ unit matrix,
and
$\bar\psi:=-i\psi^\dagger  {\hat \gamma}^0$
is the Dirac adjoint of $\psi$.
Varying the action \eqref{dirac} with respect to $\bar\psi$
yields the usual Dirac equation 
$\left(
{\hat\gamma}^a
\partial_a
-m
\right)\psi
=0$.
Assuming that
the solution is given by the Fourier form
$\psi = \int d{\hat p}\, u({\hat p})e^{i{\hat p}_a X^a}$,
where ${\hat p}_a=(-{\hat E}, \vec{\hat p})$
denotes the four-momentum
with ${\hat E}$ and $\vec{\hat p}$ being the energy and linear momentum,
respectively,
each partial mode function $u(p)$ satisfies
$\left(
i{\hat\gamma}^a
{\hat p}_a
-m
\right)
u(p)
=0$.
Acting ${\hat\gamma}^a \partial_a+m$
from the left side leads to 
$
0
=\left[
-\frac{1}{2}
\left\{{\hat \gamma}^a, {\hat \gamma}^b
\right\}
{\hat p}_a {\hat p}_b
-m^2
\right]
u({\hat p})
=
\left[
-\eta^{ab}
 {\hat p}_a {\hat p}_b
-m^2
\right]
u({\hat p})
=
\left[
{\hat E}^2
-
|\vec{\hat p}|^2
-m^2
\right]
u({\hat p})$,
and hence we obtain the standard dispersion relation
${\hat E}^2=|\vec{\hat p}|^2+m^2$.

We then consider the theory of the Dirac spinor field
 in the four-dimensional Minkowski spacetime
given by   
\begin{eqnarray}
\label{tachyonic_flat}
S_{(\psi)}
=
 \int d^4X
\left[
\frac{1}{2}
\left(
{\bar \psi}
{\hat \gamma}^5
{\hat\gamma}^a
\partial_a
\psi
-
\partial_a
{\bar \psi}
{\hat \gamma}^5
{\hat\gamma}^a 
\psi  
\right)
-M
{\bar \psi}
\psi
\right],
\end{eqnarray} 
where we have defined
\begin{eqnarray}
\label{g5}
{\hat \gamma}^5
:=
i {\hat\gamma}^0
{\hat \gamma}^1
{\hat\gamma}^2
{\hat \gamma}^3,
\qquad
\left\{{\hat \gamma}^a,{\hat\gamma}^5\right\}=0,
\qquad
{\hat \gamma}^5\cdot{\hat \gamma}^5=I_{4\times 4},
\end{eqnarray}
and $M$ is a constant.
\footnote{
The alternative theory which gives the same equation of motion for $\psi$
 is
given by  
\begin{eqnarray}
\label{tachyonic_flat_alt}
S_{(\psi)}
=
 \int d^4X
\left[
\frac{1}{2}
\left(
{\bar \psi}
{\hat\gamma}^a
\partial_a
\psi
-
\partial_a
{\bar \psi}
{\hat\gamma}^a 
\psi  
\right)
-M
{\bar \psi}
{\hat \gamma}^5
\psi
\right].
\end{eqnarray} 
As argued in Refs. \cite{Jentschura:2011ga,Ramazanoglu:2018hwk},
however, 
the variation of Eq. \eqref{tachyonic_flat_alt}
leads to the inconsistent equation for ${\bar \psi}$
(see Appendix \ref{app_a}).}
Variation of Eq. \eqref{tachyonic_flat}
yields the modified Dirac equation 
\begin{eqnarray}
\label{Dirac_tachyonic}
\left(
{\hat\gamma}^5
{\hat\gamma}^a
\partial_a
-M
\right)\psi
=0.
\end{eqnarray}
Assuming the form of the solution 
$\psi = \int d{\hat p} \, v({\hat p})e^{i{\hat p}_a X^a}$,
each mode satisfies
$\left(
i
{\hat\gamma}^a
{\hat p}_a
-M
{\hat \gamma}^5
\right)
v({\hat p})
=0.$
Acting
$
{\hat\gamma}^a
\partial_a
-M{\hat\gamma}^5$
from the left side  
results in 
$0
=
\left[
-{\hat \gamma}^a {\hat \gamma}^b
 {\hat p}_a {\hat p}_b
+
i
M
\left\{
{\hat \gamma}^a,{\hat \gamma}^5
\right\} 
{\hat p}_a
+M^2
\right]
v({\hat p})
=
\left[
-\frac{1}{2}
\left\{{\hat \gamma}^a, {\hat \gamma}^b
\right\}
 {\hat p}_a 
 {\hat p}_b
+M^2
\right]
v({\hat p})
=
\left[
-\eta^{ab}
{\hat p}_a 
{\hat p}_b
+M^2
\right]
v(\hat{p})
=
\left[
{\hat E}^2
-|\vec{\hat p}|^2
+M^2
\right]
v({\hat p})$,
and hence the dispersion relation is modified as 
${\hat E}^2  =|\vec{\hat p}|^2-M^2$.
Thus,
the modes of  $|\vec{\hat p}|<M$ suffer from the tachyonic instability.
If the constant $M$ is promoted to a position-dependent function $M(x^\mu)$,
one may be able to make the modes tachyonic only locally,
for instance, in the vicinity of black holes and compact stars.
This is the basic idea behind spontaneous spinorization \cite{Ramazanoglu:2018hwk}.

We will consider the model with the conformal coupling of the Dirac spinor field  to matter 
(see Eqs. \eqref {einstein_dirac_action}-\eqref{conformal}),
and 
study the behavior of the Dirac spinor field 
inside the high density compact stars.
We will show that 
the modified kinetic term by the insertion of ${\hat \gamma}^5$
\eqref{kin}
and the conformal coupling  \eqref{conformal}
would lead to the tachyonic growth of the Dirac spinor field 
inside the high density compact stellar backgrounds,
where the parameter $M$ is replaced 
by the trace of the energy-momentum tensor of the matter ${T}^{(m) \mu}{}_\mu$.
By the analysis in the constant density stellar backgrounds, 
we will show that this coupling  gives rise to the nontrivial profiles 
of the Dirac spinor field with any number of nodes.
We will also show that 
up to the quadratic order of the Dirac spinor fields and their adjoints
all the components of the effective energy-momentum tensor
of the Dirac field vanish
in any static and spherically symmetric stellar backgrounds.
Thus, even if the amplitude of the Dirac spinor field grows with time, 
it will not affect the spacetime geometry, indicating 
that spontaneous spinorization happens as a stealth process.
\footnote{
The term ``stealth'' has been used 
to describe the Schwarzschild/ Kerr black hole solutions
in the scalar-tensor or vector-tensor theories
where the scalar or vector field with the nontrivial profile
does not backreact on the spacetime geometry
(see e.g., Refs.~\cite{Babichev:2013cya,Chagoya:2016aar,Minamitsuji:2018vuw,Takahashi:2020hso}).
This happens 
in the case that
all the components of the effective energy-momentum tensor of the scalar or vector field 
automatically vanish
in the right-hand side of the gravitational equations of motion.}
We will dub this {\it stealth spontaneous spinorization}.
Since the vanishing effective energy-momentum tensor holds
in any static and spherically background,
stealth spontaneous spinorization would happen for any equation of state of matter. 
The tachyonic growth of the Dirac spinor field  may be quenched by nonlinear effects,
and would not affect the spacetime geometry around relativistic stars,
and not leave any observable consequences.

This paper is organized as follows:
in Sec. \ref{sec2},
we will introduce the model for spontaneous spinorization
with the coupling of the Dirac spinor field to the matter.
In Sec. \ref{sec3},
we will review the general properties of the Dirac spinor field 
in the static and spherically symmetric spacetimes.
In Sec. \ref{sec4},
we will analyze the Dirac spinor field s
in the constant density stellar backgrounds.
In Sec. \ref{sec5},
we will show 
that the effective energy-momentum tensor of the Dirac fields
vanish,
and discuss the implications for spontaneous spinorization.
The last Sec. \ref{sec6}
will be devoted to giving a brief summary and conclusion.

\section{Model}
\label{sec2}


We consider the theory
composed of the metric tensor $g_{\mu\nu}$ 
and the Dirac spinor field $\psi$,
\begin{eqnarray}
\label{einstein_dirac_action}
S
&=&
\int d^4x
\left[
\sqrt{-g}
\left(
\frac{R}{2\kappa^2}
+L_{(\psi)}
\right)
+
\sqrt{-{\tilde g}}
 L_{(m)}
\left[
{\tilde g}_{\mu\nu},
\Psi
\right]
\right],
\end{eqnarray}
with the modified kinetic term
\begin{eqnarray}
\label{kin}
{L}_{(\psi)}
:=
\frac{1}{2}
\left[
{\bar \psi}
{\hat \gamma}^5
{\gamma}^\mu 
D_\mu
\psi
-
\left(
D_\mu
{\bar \psi}
\right)
{\hat \gamma}^5
{\gamma}^\mu 
\psi  
\right],
\end{eqnarray}
where
the induces $\mu,\nu,\cdots$ run the physical spacetime with the metric $g_{\mu\nu}$, 
$\kappa^2=8\pi G/c^4$
with the gravitational constant $G$ and the speed of light $c$
(we will set $c=1$ unless necessary),
$R$ is the scalar curvature associated with $g_{\mu\nu}$, 
$\bar{\psi}:= -i \psi^\dagger {\hat\gamma}^0$
is the Dirac adjoint of $\psi$,
${\hat \gamma}^a$ denotes the Dirac gamma matrices
satisfying the anticommutation relations Eq. \eqref{dmatrix},
$\gamma^\mu:= e^\mu_a {\hat \gamma}^a$
with $e^\mu_a$ being the tetrad field satisfying 
$g_{\mu\nu} e^\mu_a e^\nu_b=\eta_{ab}$,
$D_\mu\psi:=
\partial_\mu \psi
+
\Gamma_\mu
\psi$
and 
$D_\mu {\bar\psi}
:=
\partial_\mu {\bar \psi}
-
{\bar\psi}
\Gamma_\mu$
denote the covariant derivatives
with the spin connection
$
\Gamma_\mu
:=
(1/8)
\left[
{\hat \gamma}^a,
{\hat \gamma}^b
\right]
e_a^\nu \nabla_\mu e_{b\nu}$,
${\tilde g}_{\mu\nu}$
represents the Jordan frame metric
related to the Einstein frame metric $g_{\mu\nu}$ by 
\begin{eqnarray}
\label{conformal}
{\tilde g}_{\mu\nu}
=
e^{F
\left(
\Phi
\right)}
g_{\mu\nu},
\end{eqnarray}
with $\Phi:= {\bar \psi}
\psi$,
and 
$\Psi$ is the matter field,
respectively.

Varying the action \eqref{einstein_dirac_action}
with respect to the metric $g_{\mu\nu}$
yields the gravitational equation of motion
\begin{eqnarray}
\label{eq_model_a}
G_{\mu\nu}
=\kappa^2
\left(
T^{(\psi)}_{\mu\nu}
+
e^{F}
{\tilde T}^{(m)}_{\mu\nu}
\right),
\end{eqnarray}
where we have defined 
the energy-momentum tensor for modified kinetic term \eqref{kin} 
\begin{eqnarray}
T^{(\psi)}_{\mu\nu}
:=
-\frac{2}{\sqrt{-g}}
  \frac{\delta \left(\sqrt{-g} L_{(\psi)}\right) }{\delta g^{\mu\nu}}
=
-
\frac{1}{2}
\left[
{\bar \psi}
e_{b(\mu}
{\hat \gamma}^5
{\hat\gamma}^b
D_{\nu)}
\psi
-
\left(
D_{(\mu}
{\bar \psi}
\right)
{\hat \gamma}^5
{\hat\gamma}^b
e_{b |\nu)}
\psi  
\right],
\end{eqnarray}
and the energy-momentum tensor of the matter defined in the Jordan frame
\begin{eqnarray}
{\tilde T}^{(m)}_{\mu\nu}
:=
-\frac{2}{\sqrt{-{\tilde g}}}
\frac{
\delta 
\left(
\sqrt{-{\tilde g}}
{L}_{(m)} 
 \right)
}
{\delta {\tilde g}^{\mu\nu}}.
\end{eqnarray}
Here we employed
$\delta \left(\sqrt{-{\tilde g}} L_{(m)} \right)/\delta g^{\mu\nu}
=$
$\left(\delta {\tilde g}^{\alpha\beta}/ \delta {g}^{\mu\nu}\right)
\left(
\delta \left(\sqrt{-{\tilde g}} L_{(m)} \right)/\delta {\tilde g}^{\alpha\beta}
\right)
=$
$-
\left(
e^{-F}\sqrt{-\tilde g}/2
\right)$
\\
$\times$
$\left[
-\left(2/\sqrt{-\tilde g}\right)
\delta \left(\sqrt{-{\tilde g}} L_{(m)} \right)/
\delta {\tilde g}^{\mu\nu}
\right]$
$=-\left(e^F \sqrt{-g}/2\right)
{\tilde T}^{(m)}_{\mu\nu}$.

On the other hand, 
varying the action with respect to ${\bar\psi}$
and multiplying ${\hat \gamma}^5$ from the left side yields the Dirac equation
\begin{eqnarray}
\label{eq_model_b}
\left(
{\gamma}^\mu 
D_\mu 
+
\frac{e^{2F} F_\Phi}{2}
{\hat \gamma}^5
{\tilde T}^{(m)\mu}{}_\mu
\right)
\psi
=0,
\end{eqnarray}
where $F_\Phi:= \partial F/\partial \Phi$.

We further assume that $F(\Phi)$ is the regular function of $\Phi$
which can be expanded as 
\begin{eqnarray}
\label{coupling}
F(\Phi):= 1+ \beta_1 \Phi+{\cal O} (\Phi^2),
\end{eqnarray}
where $\beta_1$ denotes the dimensionful coupling constant.
In the limit of the small amplitude,
the leading contributions to
Eqs. \eqref{eq_model_a} and \eqref{eq_model_b}
are given by 
\begin{eqnarray}
\label{Dirac_general relativity}
&&
G_{\mu\nu}
-\kappa^2
{T}^{(m)}_{\mu\nu}
+
{\cal O} (\psi^2)
=0,
\\
\label{Dirac_inside}
&&
\left(
{\gamma}^\mu 
D_\mu 
+
\frac{\beta_1}{2}
{\hat \gamma}^5
{T}^{(m)\mu}{}_\mu
\right)
\psi
+
{\cal O} (\psi^3)
=0,
\end{eqnarray}
respectively,
where 
we have defined the matter energy-momentum tensor  in the Einstein frame
${T}^{(m)}_{\mu\nu}
:= 
-
\left(
2/\sqrt{-g}\right)$\\
$\times \delta 
\left(
\sqrt{-{\tilde g}} {L}_{(m)} 
 \right)
/\delta {g}^{\mu\nu}$,
and 
${\cal O} (\psi^2)$ and ${\cal O} (\psi^3)$
represent the quadratic and cubic order combinations of 
$\psi$ and ${\bar \psi}$ and their first order derivatives, respectively.
We note that
at $\psi=0$
the matter energy-momentum tensor in the Jordan frame ${\tilde T}^{(m)}_{\mu\nu}$
coincides 
with that in the Einstein frame $T^{(m)}_{\mu\nu}$.
Thus, 
at the leading order
Eq. \eqref{Dirac_general relativity}
provides the metric solution $g_{\mu\nu}$ in general relativity
and 
Eq. \eqref{Dirac_inside}
describes the propagation of $\psi$
on top of the general relativistic backgrounds.
We also observe
that in the linearized theory 
the term $(\beta_1/2){T}^{(m)\mu}{}_\mu$
in Eq. \eqref{Dirac_inside}
plays the same role as the tachyonic mass term $M$
in Eq. \eqref{Dirac_tachyonic}
only in the regions
where $|{T}^{(m)\mu}{}_\mu|$ is large.

\section{Dirac spinor fields in static and spherically symmetric spacetimes}
\label{sec3}

\subsection{Static and spherically symmetric spacetimes}
\label{sec31}

We consider the static and spherically symmetric spacetime,
where the Einstein frame metric is given by 
\begin{eqnarray}
\label{sss}
ds^2
:=
g_{\mu\nu}dx^\mu dx^\nu
=
-e^{\nu}dt^2
+e^{\lambda}dr^2
+r^2
\left(
d\theta^2
+\sin^2\theta\, d\phi^2
\right),
\end{eqnarray}
where
$t$, $r$, $(\theta, \phi)$
are the temporal, radial, and spherical angular coordinates,
respectively,
$\nu(r)$ and $\lambda(r)$
are the functions of  $r$.
The nonzero components of the tetrad field are given by 
$e^t_0= e^{-\nu/2}$,
$e^r_1= e^{-\lambda/2}$,
$e^\theta_2= 1/r$,
and 
$e^\phi_3= 1/(r\sin\theta)$.
We adopt the Weyl representation of  the Dirac gamma matrices 
in the Minkowski spacetime satisfying 
${\hat \gamma}^1=i{\tilde \gamma}^3$,
${\hat \gamma}^2=i{\tilde \gamma}^1$,
${\hat \gamma}^3=i{\tilde \gamma}^2$,
and 
${\hat \gamma}^0=i{\tilde \gamma}^0$,
with 
$
{\tilde \gamma}^0
=
\begin{pmatrix}
0 & I_{2\times 2} \\
I_{2\times 2} & 0 \\
\end{pmatrix},
$
$
{\tilde \gamma}^1
=
\begin{pmatrix}
0& \sigma_1 \\
-\sigma_1 & 0 \\
\end{pmatrix}
$,
$
{\tilde \gamma}^2
=
\begin{pmatrix}
0& \sigma_2 \\
-\sigma_2 & 0 \\
\end{pmatrix},
$
and 
$
{\tilde \gamma}^3
=
\begin{pmatrix}
0& \sigma_3 \\
-\sigma_3 & 0 \\
\end{pmatrix}$,
where the Pauli matrices are given by 
${\sigma}_1
=
\begin{pmatrix}
0& 1 \\
1 & 0 \\
\end{pmatrix},$
${\sigma}_2
=
\begin{pmatrix}
0& -i \\
i & 0 \\
\end{pmatrix}$,
${\sigma}_3
=
\begin{pmatrix}
1& 0 \\
0& -1 \\
\end{pmatrix}$.
The fifth gamma matrix 
$\gamma^5$ in the curved spacetime is defined by 
$
\gamma^5
:=i \sqrt{-g}
\gamma^t
\gamma^r
\gamma^\theta
\gamma^\phi
=
\begin{pmatrix}
-I_{2\times2} & 0 \\
0 &  I_{2\times 2} \\
\end{pmatrix}
={\hat \gamma}^5$.
The components of the spin connection
are given by
$
\Gamma_t
=
-(e^{\frac{\nu-\lambda}{2}}/4)
\nu'
{\hat \gamma}^0
{\hat \gamma}^1$,
$\Gamma_r=0$,
$\Gamma_\theta
=
-(1/2)
e^{-\frac{\lambda}{2}}
{\hat \gamma}^1
{\hat \gamma}^2$,
and 
$\Gamma_\phi
=
-(1/2)
\left(
{\hat \gamma}^2
{\hat \gamma}^3
\cos\theta
+
{\hat \gamma}^1
{\hat \gamma}^3 
e^{-\frac{\lambda}{2}}
\sin\theta
\right)$.

Before analyzing our theory, 
we briefly review the static and spherically
symmetric configuration of the Dirac spinor field $\psi$
in the spacetime Eq. \eqref{sss}.

\subsection{Massive Dirac spinor fields}
\label{sec32}

First,
we consider the massive Dirac spinor field theory given by 
\begin{eqnarray}
\label{lag_massive}
S_{(\psi)}
&=&
\int d^4 x
\sqrt{-g}
L_{(\psi)}
:=
\int d^4x
\sqrt{-g}
\left[
\frac{1}{2}
\left(
{\bar \psi}
{\gamma}^\mu 
D_\mu
\psi
-
\left(
D_\mu
{\bar \psi}
\right)
{\gamma}^\mu 
\psi  
\right)
-m {\bar\psi} \psi
\right],
\end{eqnarray}
whose variation with respect to ${\bar \psi}$
gives the Dirac equation
\begin{eqnarray}
\label{massive}
\left(
{\gamma}^\mu D_\mu
-m
\right)
\psi
=0.
\end{eqnarray}
Using the properties
${\hat\gamma}^0 \Gamma_\mu
+\left( \Gamma_\mu \right)^\dagger
{\hat \gamma}^0
=0$
and
${\hat\gamma}^0 \gamma^\mu
+\left( \gamma^\mu \right)^\dagger
{\hat \gamma}^0
=0$,
the adjoint to the Dirac equation \eqref{massive}
is given by 
${\bar\psi}
\left[
\left(
\overleftarrow{\partial}_\mu
-\Gamma_\mu
\right)
{\gamma}^\mu
+m
\right]
=0$. 

In order to obtain 
the spherically symmetric solutions,
we have to take the following two Dirac spinor fields 
into consideration
at the same time
\cite{Herdeiro:2017fhv}
\begin{eqnarray}
\label{ansatz_dirac_m}
\psi_1
&=&
\frac{e^{-i\omega t+i \frac{\phi}{2}}}
        {r e^{\nu/4}}
\begin{pmatrix}
\cos\left(\frac{\theta}{2}\right)
\left[
(1+i) f(r)+ (1-i)g(r)
\right]
\\ 
i\sin\left(\frac{\theta}{2}\right)
\left[
(1-i) f(r)+ (1+i)g(r)
\right]
\\
-i
\cos\left(\frac{\theta}{2}\right)
\left[
(1-i) f(r)+ (1+i)g(r)
\right]
\\ 
-\sin\left(\frac{\theta}{2}\right)
\left[
(1+i) f(r)+ (1-i)g(r)
\right]
\end{pmatrix},
\\
\label{ansatz_dirac_m2}
\psi_2
&=&
\frac{e^{-i\omega t-i \frac{\phi}{2}}}
        {r e^{\nu/4}}
\begin{pmatrix}
i
\sin\left(\frac{\theta}{2}\right)
\left[
(1+i) f(r)+ (1-i)g(r)
\right]
\\ 
\cos\left(\frac{\theta}{2}\right)
\left[
(1-i) f(r)+ (1+i)g(r)
\right]
\\
\sin \left(\frac{\theta}{2}\right)
\left[
(1-i) f(r)+ (1+i)g(r)
\right]
\\ 
i\cos\left(\frac{\theta}{2}\right)
\left[
(1+i) f(r)+ (1-i)g(r)
\right]
\end{pmatrix}.
\end{eqnarray}
Substituting Eqs. \eqref{ansatz_dirac_m} and \eqref{ansatz_dirac_m2}
into Eq. \eqref{massive},
the complex radial mode functions $f(r)$ and $g(r)$ satisfy
\begin{eqnarray}
\label{radial}
&&
\left(
\frac{d}{dr}
+\frac{e^{\frac{\lambda}{2}}}
         {r}
\right)
 f(r)
=
- e^{\frac{\lambda}{2}-\frac{\nu}{2}} 
\left (
\omega 
- m e^{\frac{\nu}{2}}
\right)
g(r),
\qquad
\left(
\frac{d}{dr}
-
\frac{e^{\frac{\lambda}{2}}}
         {r}
\right)
 g(r)
=
 e^{\frac{\lambda}{2}-\frac{\nu}{2}} 
\left (
\omega 
+
m e^{\frac{\nu}{2}}
\right)
f(r).
\end{eqnarray}
Introducing the new variables by 
\begin{eqnarray}
\label{introduction_fg}
{\rm Re} (f)
=
\frac{1}{2}
\left(
Q_1+Q_2
\right),
\qquad 
{\rm Re} (g)
=
\frac{1}{2}
\left(
Q_4-Q_3
\right),
\qquad
{\rm Im} (f)
=
\frac{1}{2}
\left(
Q_3+Q_4
\right),
\qquad 
{\rm Im} (g)
=
\frac{1}{2}
\left(
Q_1-Q_2
\right),
\end{eqnarray}
Eqs. 
\eqref{ansatz_dirac_m}
and 
\eqref{ansatz_dirac_m2}
reduce to
\begin{eqnarray}
\psi_1
&=&
\frac{(1+i)e^{-i\omega t+i \frac{\phi}{2}}}
        {r e^{\nu/4}}
\begin{pmatrix}
\cos\left(\frac{\theta}{2}\right)
\left[
Q_1+i Q_3
\right]
\\ 
\sin\left(\frac{\theta}{2}\right)
\left[
Q_2+i Q_4
\right]
\\
-
\cos\left(\frac{\theta}{2}\right)
\left[
Q_2+ iQ_4
\right]
\\
- 
 \sin\left(\frac{\theta}{2}\right)
\left[
Q_1+i Q_3
\right]
\end{pmatrix},
\quad
\psi_2
=
\frac{(1-i)
e^{-i\omega t-i \frac{\phi}{2}}}
        {r e^{\nu/4}}
\begin{pmatrix}
-
\sin\left(\frac{\theta}{2}\right)
\left[
Q_1+iQ_3
\right]
\\ 
\cos\left(\frac{\theta}{2}\right)
\left[
Q_2+iQ_4
\right]
\\
\sin \left(\frac{\theta}{2}\right)
\left[
 Q_2+ iQ_4
\right]
\\ 
-
\cos\left(\frac{\theta}{2}\right)
\left[
Q_1+ i Q_3
\right]
\end{pmatrix},
\end{eqnarray}
and 
from Eq. \eqref{radial}
we find that  $Q_1$-$Q_4$ satisfy 
\begin{eqnarray}
&&
\frac{dQ_1}{dr}
+
\frac{e^{\frac{\lambda}{2}}}{r}Q_2
-m
 e^{\frac{\lambda}{2}}
Q_4
-
e^{\frac{\lambda-\nu}{2}}
{\rm Re}(\omega) 
Q_3
-
e^{\frac{\lambda-\nu}{2}}
{\rm Im}(\omega) 
Q_1
=0,
\nonumber
\\
&&
\frac{dQ_2}{dr}
+
\frac{e^{\frac{\lambda}{2}}}{r}
Q_1
+m 
e^{\frac{\lambda}{2}}
Q_3
+
e^{\frac{\lambda-\nu}{2}}
{\rm Re}(\omega) 
Q_4
+
e^{\frac{\lambda-\nu}{2}}
{\rm Im}(\omega) 
Q_2
=0,
\nonumber
\\
&&
\frac{dQ_3}{dr}
+
\frac{e^{\frac{\lambda}{2}}}{r}
Q_4
+
m
 e^{\frac{\lambda}{2}}
Q_2
+
e^{\frac{\lambda-\nu}{2}}
{\rm Re}(\omega) 
Q_1
-
e^{\frac{\lambda-\nu}{2}}
{\rm Im}(\omega) 
Q_3
=0,
\nonumber
\\
&&
\frac{dQ_4}{dr}
+
\frac{e^{\frac{\lambda}{2}}}{r}
Q_3
-
m
e^{\frac{\lambda}{2}}
Q_1
-
e^{\frac{\lambda-\nu}{2}}
{\rm Re}(\omega) 
Q_2
+
e^{\frac{\lambda-\nu}{2}}
{\rm Im}(\omega) 
Q_4
=0.
\end{eqnarray}
The energy-momentum tensor is given by
the combination of the above two Dirac spinor fields
\begin{eqnarray}
T^{(\psi)}_{\mu\nu}
&=&
\sum_{I=1,2}
T^{(\psi_I)}_{\mu\nu}
:=
-\sum_{I=1,2}
\frac{2}{\sqrt{-g}}
  \frac{\delta \left(\sqrt{-g} L_{\psi_I}\right) }{\delta g^{\mu\nu}}
\nonumber\\
&=&
-
\frac{1}{2}
\sum_{I=1,2}
\left[
{\bar \psi}_I
e_{b(\mu}
{\hat\gamma}^b
D_{\nu)}
\psi_I
-
\left(
D_{(\mu}
{\bar \psi}_I
\right)
{\hat\gamma}^b
e_{b |\nu)}
\psi_I
\right]
-m
g_{\mu\nu}
\sum_{I=1,2}
{\bar\psi}_I
\psi_I,
\end{eqnarray}
where $L_{(\psi_I)}$ is given by Eq. \eqref{lag_massive}
with the substitution of $\psi$ and $\bar \psi$ to $\psi_I$ and ${\bar \psi}_I$ ($I=1,2$).
After the summation of the two contributions, 
the energy density and pressure are given by 
\begin{eqnarray}
\rho_{(\psi)} c^2
&:=&
-T^{(\psi)}{}^t{}_t
=
e^{2{\rm Im} (\omega)t} 
\left[
\frac{4 e^{-\nu} {\rm Re}(\omega)}
       {r^2}
\left(
  Q_1^2
+ Q_2^2
+Q_3^2
+Q_4^2
\right)
+
\frac{8m e^{-\frac{\nu}{2}}}
         {r^2}
\left(
Q_1 Q_2
+Q_3Q_4
\right)
\right],
\nonumber
\\
p_{(\psi),r}
&:=&
T^{(\psi)}{}^r{}_r
=
e^{2{\rm Im} (\omega)t} 
\left[
\frac{4 e^{-\frac{\nu}{2}-\frac{\lambda}{2}}}
        {r^2}
\left(
  Q_3Q_1'
-Q_4 Q_2'
-Q_1 Q_3'
+Q_2 Q_4'
\right)
-
\frac{8m e^{-\frac{\nu}{2}}}
         {r^2}
\left(
Q_1 Q_2
+Q_3Q_4
\right)
\right],
\nonumber
\\
p_{(\psi),\theta}
&:=&
T^{(\psi)}{}^\theta{}_\theta
=
e^{2{\rm Im} (\omega)t} 
\left[
\frac{4 e^{-\frac{\nu}{2}}}
        {r^3}
\left(
  Q_2 Q_3
-Q_1 Q_4
\right)
-
\frac{8m e^{-\frac{\nu}{2}}}
         {r^2}
\left(
Q_1 Q_2
+Q_3Q_4
\right)
\right],
\nonumber\\
p_{(\psi),\phi}
&:=&
T^{(\psi)}{}^\phi{}_\phi
=p_{(\psi),\theta},
\end{eqnarray}
and all the off-diagonal components vanish, if 
\begin{eqnarray}
-e^{\frac{\lambda-\nu}{2}}
{\rm Re}(\omega)
\left(
  Q_1^2-Q_2^2
+Q_3^2-Q_4^2
\right)
-Q_3 Q_1'
-Q_4 Q_2'
+Q_1 Q_3'
+Q_2 Q_4'
=0.
\end{eqnarray}
We note that 
the energy-momentum tensor
for each individual field $T^{(\psi_I)}_{\mu\nu}$
is not diagonal
and depends on $\theta$ as well as $r$.
In other words, 
the two fields \eqref{ansatz_dirac_m} and \eqref{ansatz_dirac_m2}
are necessary to be compatible with the spherical symmetry.
The static and spherically symmetric solutions can be obtained
for ${\rm Im} (\omega)=0$.

In the flat Minkowski spacetime with $\nu=\lambda=0$,
one solution is obtained for $\omega=m$
with $f(r)=C_1/r$ and $g(r)=-2mC_1+r C_2$,
where $C_1$ and $C_2$ are real integration constants.
Thus, the solution behaves as an oscillatory one $\psi\sim e^{-imt}$.
The other solution is obtained for $\omega=-m$
with $f(r)=2m D_1 r^2/3+D_2/r$ and $g(r)=r D_1$,
where $D_1$ and $D_2$ are real integration constants,
which also behave as an oscillatory one $\psi\sim e^{imt}$.

\subsection{Tachyonic Dirac spinor fields}
\label{sec33}

Second,
we consider a tachyonic massive Dirac spinor field theory
in the curved spacetime
\begin{eqnarray}
\label{lag_tachyon}
S_{(\psi)}
&=&
\int 
d^4x
\sqrt{-g}
L_{(\psi)}
=
\int d^4x
\sqrt{-g}
\left(
\frac{1}{2}
\left(
{\bar \psi}
{\hat \gamma}^5
{\gamma}^\mu 
D_\mu
\psi
-
\left(
D_\mu
{\bar \psi}
\right)
{\hat \gamma}^5
{\gamma}^\mu 
\psi  
\right)
-M {\bar\psi} \psi
\right),
\end{eqnarray}
whose variation with respect to ${\bar \psi}$
gives the Dirac equation
\begin{eqnarray}
\label{tachyonic}
\left(
{\gamma}^\mu D_\mu
-M
{\hat \gamma}^5
\right)
\psi
=0.
\end{eqnarray}
Using the properties
${\hat\gamma}^0 \Gamma_\mu
+\left( \Gamma_\mu \right)^\dagger
{\hat \gamma}^0
=0$,
${\hat\gamma}^0 {\hat \gamma}^5
+ {\hat \gamma}^5 {\hat \gamma}^0
=0$,
and
${\hat\gamma}^0 \gamma^\mu
+\left( \gamma^\mu \right)^\dagger
{\hat \gamma}^0
=0$,
the adjoint to the Dirac equation \eqref{tachyonic}
is given by 
${\bar\psi}
\left[
\left(
\overleftarrow{\partial}_\mu
-\Gamma_\mu
\right)
{\gamma}^\mu
-M{\hat \gamma}^5
\right]
=0$,
or equivalently 
by multiplying ${\hat \gamma}^5$ from the left side
${\bar\psi}
\left[
\left(
\overleftarrow{\partial}_\mu
-\Gamma_\mu
\right)
{\hat \gamma}^5
{\gamma}^\mu
+M
\right]
=0$.

Similarly to the case \eqref{lag_massive}, 
to obtain the 
spherically symmetric solutions, 
we have to consider the following two Dirac fields
\cite{Herdeiro:2017fhv}:
\begin{eqnarray}
\label{ansatz_dirac_t}
\psi_1
&=&
\frac{e^{-i\omega t+i \frac{\phi}{2}}}
        {r e^{\nu/4}}
\begin{pmatrix}
\cos\left(\frac{\theta}{2}\right)
\left[
(1+i) f(r)+ (1-i)g(r)
\right]
\\ 
i\sin\left(\frac{\theta}{2}\right)
\left[
(1-i) f(r)+ (1+i)g(r)
\right]
\\
\cos\left(\frac{\theta}{2}\right)
\left[
(1-i) f(r)+ (1+i)g(r)
\right]
\\ 
-i \sin\left(\frac{\theta}{2}\right)
\left[
(1+i) f(r)+ (1-i)g(r)
\right]
\end{pmatrix},
\\
\label{ansatz_dirac2t}
\psi_2
&=&
\frac{e^{-i\omega t-i \frac{\phi}{2}}}
        {r e^{\nu/4}}
\begin{pmatrix}
i
\sin\left(\frac{\theta}{2}\right)
\left[
(1+i) f(r)+ (1-i)g(r)
\right]
\\ 
\cos\left(\frac{\theta}{2}\right)
\left[
(1-i) f(r)+ (1+i)g(r)
\right]
\\
i
\sin \left(\frac{\theta}{2}\right)
\left[
(1-i) f(r)+ (1+i)g(r)
\right]
\\ 
-\cos\left(\frac{\theta}{2}\right)
\left[
(1+i) f(r)+ (1-i)g(r)
\right]
\end{pmatrix}.
\end{eqnarray}
Compared to Eqs. \eqref{ansatz_dirac_m} and \eqref{ansatz_dirac_m2},
the third and fourth components are different 
by the multiplication of the factor $(-i)$.
Substituting Eqs. \eqref{ansatz_dirac_t} and \eqref{ansatz_dirac2t}
into Eq. \eqref{tachyonic},
we find that 
the complex radial mode functions $f(r)$ and $g(r)$ satisfy
\begin{eqnarray}
&&
\label{tachyon_fg}
\left(
\frac{d}{dr}
+\frac{e^{\frac{\lambda}{2}}}
         {r}
\right)
 f(r)
=
- e^{\frac{\lambda}{2}-\frac{\nu}{2}} 
\left (
\omega 
-i M e^{\frac{\nu}{2}}
\right)
g(r),
\qquad
\left(
\frac{d}{dr}
-
\frac{e^{\frac{\lambda}{2}}}
         {r}
\right)
 g(r)
=
 e^{\frac{\lambda}{2}-\frac{\nu}{2}} 
\left (
\omega 
+
iM e^{\frac{\nu}{2}}
\right)
f(r).
\end{eqnarray}
Introducing the new variables $Q_1$-$Q_4$
as Eq. \eqref{introduction_fg},
the two fields 
\eqref{ansatz_dirac_t}
and 
\eqref{ansatz_dirac2t}
reduce to
\begin{eqnarray}
\psi_1
&=&
\frac{(1+i)e^{-i\omega t+i \frac{\phi}{2}}}
        {r e^{\nu/4}}
\begin{pmatrix}
\cos\left(\frac{\theta}{2}\right)
\left[
Q_1+i Q_3
\right]
\\ 
\sin\left(\frac{\theta}{2}\right)
\left[
Q_2+i Q_4
\right]
\\
\cos\left(\frac{\theta}{2}\right)
\left[
-iQ_2+Q_4
\right]
\\ 
 \sin\left(\frac{\theta}{2}\right)
\left[
-iQ_1+ Q_3
\right]
\end{pmatrix},
\quad
\psi_2
=
\frac{(1-i)
e^{-i\omega t-i \frac{\phi}{2}}}
        {r e^{\nu/4}}
\begin{pmatrix}
-
\sin\left(\frac{\theta}{2}\right)
\left[
Q_1+iQ_3
\right]
\\ 
\cos\left(\frac{\theta}{2}\right)
\left[
Q_2+iQ_4
\right]
\\
-
\sin \left(\frac{\theta}{2}\right)
\left[
-i Q_2+Q_4
\right]
\\ 
\cos\left(\frac{\theta}{2}\right)
\left[
-iQ_1+Q_3
\right]
\end{pmatrix},
\end{eqnarray}
and 
from Eq. \eqref{tachyon_fg}
we find that  $Q_1$-$Q_4$
satisfy 
\begin{eqnarray}
&&
\label{tac1}
\frac{dQ_1}{dr}
+
\frac{e^{\frac{\lambda}{2}}}{r}
\left(
1
-Mr
\right)
Q_2
-
e^{\frac{\lambda-\nu}{2}}
{\rm Re}(\omega) 
Q_3
-
e^{\frac{\lambda-\nu}{2}}
{\rm Im}(\omega) 
Q_1
=0,
\nonumber
\\
\label{tac2}
&&
\frac{dQ_2}{dr}
+
\frac{e^{\frac{\lambda}{2}}}{r}
\left(
1
+Mr 
\right)
Q_1
+
e^{\frac{\lambda-\nu}{2}}
{\rm Re}(\omega) 
Q_4
+
e^{\frac{\lambda-\nu}{2}}
{\rm Im}(\omega) 
Q_2
=0,
\nonumber
\\
&&
\frac{dQ_3}{dr}
+
\frac{e^{\frac{\lambda}{2}}}{r}
\left(
1
-Mr
\right)
Q_4
+
e^{\frac{\lambda-\nu}{2}}
{\rm Re}(\omega) 
Q_1
-
e^{\frac{\lambda-\nu}{2}}
{\rm Im}(\omega) 
Q_3
=0,
\nonumber
\\
&&
\frac{dQ_4}{dr}
+
\frac{e^{\frac{\lambda}{2}}}{r}
\left(
1
+Mr
\right)
Q_3
-
e^{\frac{\lambda-\nu}{2}}
{\rm Re}(\omega) 
Q_2
+
e^{\frac{\lambda-\nu}{2}}
{\rm Im}(\omega) 
Q_4
=0.
\end{eqnarray}
Similarly, 
the energy-momentum tensor is given by
the summation of the contributions of the above two fields
\begin{eqnarray}
T^{(\psi)}_{\mu\nu}
=
\sum_{I=1,2}
T^{(\psi_I)}_{\mu\nu}
&=&
-
\frac{1}{2}
\sum_{I=1,2}
\left[
{\bar \psi}_I
{\hat \gamma}^5
e_{b(\mu}
{\hat\gamma}^b
D_{\nu)}
\psi_I
-
\left(
D_{(\mu}
{\bar \psi}_I
\right)
{\hat \gamma}^5
{\hat\gamma}^b
e_{b |\nu)}
\psi_I
\right]
-M 
g_{\mu\nu}
\sum_{I=1,2}
{\bar\psi}_I
\psi_I.
\end{eqnarray}
The contribution from each individual Dirac field is, 
respectively,
given by
\begin{eqnarray}
\label{38}
\rho_{(\psi_1)} c^2
&=&
e^{2{\rm Im}(\omega) t}
\left[ 
\frac{
2e^{-\nu} 
{\rm Re}(\omega)}{r^2}\cos\theta
\left(
Q_1^2-Q_2^2
+Q_3^2-Q_4^2
\right)
-
\frac{4Me^{-\frac{\nu}{2}}}
         {r^2}
\left(
  Q_2 Q_3
-Q_1Q_4
\right)
\cos\theta
\right],
\nonumber
\\
 p_{(\psi_1),r}
&=&
e^{2{\rm Im}(\omega) t}
\left[
\frac{2e^{-\frac{\lambda+\nu}{2}}}{r^2}
\cos\theta
\left(
 Q_3 Q_1'
+Q_4 Q_2'
-Q_1Q_3'
-Q_2Q_4'
\right)
+
\frac{4Me^{-\frac{\nu}{2}}}
         {r^2}
\left(
  Q_2 Q_3
-Q_1Q_4
\right)
\cos\theta
\right],
\nonumber\\
 p_{(\psi_1),\theta}
&=&
p_{(\psi_1),\phi}
=0,
\end{eqnarray}
and
$\rho_{(\psi_2)} 
=
-\rho_{(\psi_1)}$,
$p_{(\psi_2),r}
=- p_{(\psi_1),r}$,
$ p_{(\psi_2),\theta}
=p_{(\psi_2),\phi}
=0$.
We note that
each of  $T^{(\psi_I)}_{\mu\nu}$ ($I=1,2$)
is not diagonal
and 
the diagonal components depend on $\theta$
as well as $r$.
However, 
after their summation,
we confirm that 
all the off-diagonal components vanish
and the diagonal components become
\begin{eqnarray}
\rho_{(\psi)} c^2
=
p_{(\psi),r}
=
p_{(\psi),\theta}
=
p_{(\psi),\phi}
=
0.
\end{eqnarray}
Thus, 
the tachyonic Dirac spinor field
possesses the vanishing energy-momentum tensor
in the entire spacetime.
We note that 
this property of the vanishing energy-momentum tensor 
holds 
for any complex number of $\omega$,
and 
even for the case that 
the time-dependent factor $e^{-i\omega t}$
in Eqs.~\eqref{ansatz_dirac_t} and \eqref{ansatz_dirac2t}
is replaced
by any complex function of time $h(t)$.
We also note that since $\rho_{(\psi_2)} =-\rho_{(\psi_1)}$ as shown below Eq. \eqref{38},
one of the two Dirac fields has the positive energy density,
while the other has the negative one.
However,  the negative energy density of one field does not  imply any pathology,
as only the combination of the two Dirac fields has the
physical meaning to maintain the static and spherically symmetric 
spacetime. 
In the model \eqref{einstein_dirac_action} with Eq. \eqref{kin}, 
where the constant parameter $M$
is promoted to a position-dependent function $M(r)$,
the argument here holds at the quadratic order of $\psi$ and ${\bar\psi}$.

In the Minkowski spacetime $\nu=\lambda=0$,
one solution is obtained for $\omega=iM$
with $f(r)=C_1/r$ and $g(r)=-2i MC_1+r C_2$,
where $C_1$ and $C_2$ are complex integration constants,
which behaves as an exponentially growing one $\psi\sim e^{Mt }$.
The other solution is obtained for $\omega=-iM$
with $f(r)=2iM D_1 r^2/3+D_2/r$ and $g(r)=r D_1$,
where $D_1$ and $D_2$ are complex integration constants,
which also behave as the decaying solution with time $\psi\sim e^{-Mt}$.

\section{Dirac spinor fields in the constant density stellar backgrounds}
\label{sec4}

\subsection{The constant density star solution in general relativity}
\label{sec41}

We then consider a star composed of 
an  incompressible fluid, ${\tilde \rho}=\rho_0={\rm const}$
in general relativity.
The interior metric and pressure ($r<\mathcal{R}$) are then given by 
\begin{eqnarray}
\label{gr_exa}
e^{\lambda(r)}
&=
&\left(1-\frac{2GM_0 r^2}{c^2\mathcal{R}^3}\right)^{-1},
\qquad
e^{\nu(r)}
=
c^2
\left[
\frac{3}{2}
 \left(1-\frac{2GM_0}{c^2\mathcal{R}}\right)^{1/2}
-\frac{1}{2}
\left(1-\frac{2GM_0 r^2}{c^2 \mathcal{R}^3}\right)^{1/2}
\right]^2,
\nonumber\\
{\tilde p}(r)&=&\rho_0  c^2
\frac{\left(1-\frac{2GM_0 r^2}{c^2 \mathcal{R}^3}\right)^{1/2}
      -\left(1-\frac{2GM_0}{c^2 \mathcal{R}}\right)^{1/2}}
      {3\left(1-\frac{2GM_0}{c^2\mathcal{R}}\right)^{1/2}
        -\left(1-\frac{2GM_0 r^2}{c^2\mathcal{R}^3}\right)^{1/2}},
\end{eqnarray}
where at the surface of the star $r=\mathcal{R}$, ${\tilde p}(\mathcal{R})=0$
and $M_0$ and ${\cal C}$ are the total mass and compactness,
$M_0:=4\pi \mathcal{R}^3\rho_0 /3$
and 
${\cal C}:=GM_0/(c^2\mathcal{R})$,
respectively.
On the other hand, 
the spacetime exterior to the constant density star ($r>\mathcal{R}$)
is given by the Schwarzschild metric 
\begin{eqnarray}
\label{external_metric}
e^{\lambda(r)}
&=&
e^{-\nu(r)} c^2
=\left(1-\frac{2GM_0}{c^2r}\right)^{-1}.
\end{eqnarray}
For the coupling function \eqref{coupling},
the Dirac equation \eqref{Dirac_inside}
inside the star
reduces to 
 \begin{eqnarray}
\label{historia}
\left(
{\gamma}^\mu 
D_\mu 
+
\frac{\beta_1}{2}
{\hat \gamma}^5
\left(
-\rho_0c^2 +3{\tilde p}(r)
\right)
\right)
\psi
+
{\cal O}
(\psi^3)
=0,
\end{eqnarray}
while outside the star $\gamma^\mu D_\mu\psi+{\cal O} (\psi^3)=0$.
Since the energy density
has a discontinuity from $\rho_0 c^2$ to $0$
across the surface of the star
from the inside to the outside,
the second term in Eq. \eqref{historia}
has a discontinuity across it,
and consequently
the first order derivative $\partial_r \psi$
should also have a discontinuity.
We note that 
such a sudden discontinuity of $\partial_r\psi$ is somewhat an artifact
of the idealization of the background solutions,
and
since 
in realistic relativistic stellar backgrounds
the matter energy density would quickly but smoothly go to zero,
the discontinuity of $\partial_r \psi$ would also be smoothened.

The simplest solution to Eq. \eqref{historia}
is the trivial solution $\psi=0$,
which corresponds to the general relativistic solution.
As in the case of spontaneous scalarization,
in the mechanism of spontaneous spinorization
the tachyonic instability of the trivial solution $\psi=0$
should lead to the nontrivial profile 
of the Dirac field  $\psi=\psi (x^\mu)$.
In the next section,
we will discuss the configuration of 
the nontrivial solutions
which are expected to be the end point of the tachyonic instability.

\subsection{The interior solution of the Dirac spinor fields}
\label{sec42}

As in Sec.~\ref{sec42}
following Ref.  \cite{Herdeiro:2017fhv},
we have to take 
the two Dirac spinors 
into consideration
\begin{eqnarray}
\label{internal_dirac_p}
\psi_1
=
\psi_{1,{\rm in}}
&=&
\frac{e^{-i\omega t+ i \frac{\phi}{2}}}
        {r e^{\nu/4}}
\begin{pmatrix}
\cos\left(\frac{\theta}{2}\right)
\left[
(1+i) f_{\rm in} (r)+ (1-i)g_{\rm in}(r)
\right]
\\ 
i\sin\left(\frac{\theta}{2}\right)
\left[
(1-i) f_{\rm in}(r)+ (1+i)g_{\rm in}(r)
\right]
\\
\cos\left(\frac{\theta}{2}\right)
\left[
(1-i) f_{\rm in}(r)+ (1+i)g_{\rm in}(r)
\right]
\\ 
-i \sin\left(\frac{\theta}{2}\right)
\left[
(1+i) f_{\rm in}(r)+ (1-i)g_{\rm in}(r)
\right]
\end{pmatrix},
\\
\label{internal_dirac_m}
\psi_2
=
\psi_{2,{\rm in}}
&=&
\frac{e^{-i\omega t -i \frac{\phi}{2}}}
        {r e^{\nu/4}}
\begin{pmatrix}
i
\sin\left(\frac{\theta}{2}\right)
\left[
(1+i) f_{\rm in}(r)+ (1-i)g_{\rm in}(r)
\right]
\\ 
\cos\left(\frac{\theta}{2}\right)
\left[
(1-i) f_{\rm in}(r)+ (1+i)g_{\rm in}(r)
\right]
\\
i
\sin \left(\frac{\theta}{2}\right)
\left[
(1-i) f_{\rm in}(r)+ (1+i)g_{\rm in}(r)
\right]
\\ 
-\cos\left(\frac{\theta}{2}\right)
\left[
(1+i) f_{\rm in}(r)+ (1-i)g_{\rm in}(r)
\right]
\end{pmatrix},
\end{eqnarray}
where $\omega$ is the real frequency.
Substituting 
Eqs. \eqref{internal_dirac_p} and \eqref{internal_dirac_m}
into Eq. \eqref{historia},
we find 
the radial mode functions satisfy
\begin{eqnarray}
\label{qeq1}
&&
\frac{dQ_{1,{\rm in}}}
       {dr}
+
\frac{e^{\frac{\lambda}{2}}}{2r}
\left(
2
+\beta_1 r 
{T}^{(m)\mu}{}_\mu
\right)
Q_{2,{\rm in}}
-
e^{\frac{\lambda-\nu}{2}}
\omega
Q_{3,{\rm in}}
=0,
\,\,
\frac{dQ_{2,{\rm in}}}
        {dr}
+
\frac{e^{\frac{\lambda}{2}}}{2r}
\left(
2
-\beta_1 r 
{T}^{(m)\mu}{}_\mu
\right)
Q_{1,{\rm in}}
+
e^{\frac{\lambda-\nu}{2}}
\omega
Q_{4,{\rm in}}
=0,
\nonumber
\\
&&
\frac{dQ_{3,{\rm in}}}
        {dr}
+
\frac{e^{\frac{\lambda}{2}}}{2r}
\left(
2
+\beta_1 r {T}^{(m)\mu}{}_\mu
\right)
Q_{4,{\rm in}}
+
e^{\frac{\lambda-\nu}{2}}
\omega
Q_{1,{\rm in}}
=0,
\,\,
\frac{dQ_{4,{\rm in}}}{dr}
+
\frac{e^{\frac{\lambda}{2}}}{2r}
\left(
2
-\beta_1 r {T}^{(m)\mu}{}_\mu
\right)
Q_{3,{\rm in}}
-
e^{\frac{\lambda-\nu}{2}}
\omega
Q_{2,{\rm in}}
=0,
\end{eqnarray}
where
\begin{eqnarray}
\label{fg_in}
&&
{\rm Re} (f_{\rm in})
=
\frac{1}{2}
\left(
Q_{1,{\rm in}}+Q_{2,{\rm in}}
\right),
\,\,
{\rm Re} (g_{\rm in})
=
\frac{1}{2}
\left(
Q_{4,{\rm in}}-Q_{3,{\rm in}}
\right),
\nonumber\\
&&
{\rm Im} (f_{\rm in})
=
\frac{1}{2}
\left(
Q_{3,{\rm in}}
+Q_{4,{\rm in}}
\right),
\,\,
{\rm Im} (g_{\rm in})
=
\frac{1}{2}
\left(
Q_{1,{\rm in}}
-Q_{2,{\rm in}}
\right).
\end{eqnarray}
In the vicinity of the center of the star, $r=0$,
the interior solutions satisfying the regularity boundary conditions
are given by
\begin{eqnarray}
Q_{1,{\rm in}}=  C_{1,{\rm in}} r
                    + {\cal O} (r^2),
\quad  
Q_{2,{\rm in}}= - C_{1,{\rm in}} r+ {\cal O} (r^2),
\quad 
Q_{3,{\rm in}}=  C_{3,{\rm in}} r+ {\cal O} (r^2),
\quad  
Q_{4,{\rm in}}= - C_{3,{\rm in}} r+ {\cal O} (r^2),
\end{eqnarray}
where $C_{1,{\rm in}}$ and $C_{3,{\rm in}}$ are constants,
which represent the two physically independent modes
in the static and spherically symmetric spacetime.

We note that 
if $Q_{k,{\rm in}}(r)$ ($k=1, 2, 3, 4$) is a solution,
then $c' Q_{k,{\rm in}} (r)$ with constant $c'$
is also a solution.
For $\omega=0$,
according to
the structure of Eq. \eqref{qeq1},
once the coupled equations for $Q_{1,{\rm in}}$ and $Q_{2,{\rm in}}$
can be solved
under the regularity boundary conditions at the center of the star,
those for $Q_{3,{\rm in}}$ and $Q_{4,{\rm in}}$
can also be automatically solved
under the same boundary conditions
with the following properties $Q_{3,{\rm in}}/Q_{1,{\rm in}}=Q_{4,{\rm in}}/Q_{2,{\rm in}}$.
Due to the properties of the external solutions (see Sec. \ref{sec4d}),
for the numerical analysis we have to focus on $\omega=0$,
and hence the subsystem of $Q_{1,{\rm in}}$ and $Q_{2,{\rm in}}$.

\subsection{The exterior solution of the Dirac spinor fields}
\label{sec4d}

The two fields inside the star
\eqref{internal_dirac_p} 
and 
\eqref{internal_dirac_m} 
can be matched to 
those outside the star,
$\psi_{1,{\rm out}}$ and $\psi_{2,{\rm out}}$,
given by \eqref{internal_dirac_p}
with the replacement 
of the subscripts ``${\rm in}$'' with ``${\rm out}$.''
The radial mode functions
$Q_{1,{\rm out}}$,
$Q_{2,{\rm out}}$,
$Q_{3,{\rm out}}$,
and 
$Q_{4,{\rm out}}$,
given by Eq. \eqref{fg_in}
with ${\rm in}$  $\to$ ${\rm out}$,
obey
\begin{eqnarray}
&&
\frac{dQ_{1,{\rm out}}}
       {dr}
+
\frac{e^{\frac{\lambda}{2}}}{r}
Q_{2,{\rm out}}
-
e^{\frac{\lambda-\nu}{2}}
\omega
Q_{3,{\rm out}}
=0,
\quad
\frac{dQ_{2,{\rm out}}}
        {dr}
+
\frac{e^{\frac{\lambda}{2}}}{r}
Q_{1,{\rm out}}
+
e^{\frac{\lambda-\nu}{2}}
\omega
Q_{4,{\rm out}}
=0,
\nonumber
\\
&&
\frac{dQ_{3,{\rm out}}}
        {dr}
+
\frac{e^{\frac{\lambda}{2}}}{r}
Q_{4,{\rm out}}
+
e^{\frac{\lambda-\nu}{2}}
\omega
Q_{1,{\rm out}}
=0,
\quad
\frac{dQ_{4,{\rm out}}}{dr}
+
\frac{e^{\frac{\lambda}{2}}}{r}
Q_{3,{\rm out}}
-
e^{\frac{\lambda-\nu}{2}}
\omega
Q_{2,{\rm out}}
=0.
\end{eqnarray}
For $\omega\neq 0$,
we have numerically confirmed
that in the Minkowski spacetime
the solutions are oscillatory 
in the entire spacetime.
Even under the presence of the star
we expect the similar oscillatory behaviors
toward the spatial infinity.
Thus,
it is hard to imagine 
that such oscillatory solutions arise
from the tachyonic instability of the trivial solution $\psi=0$, 
and only the solution obeying  the boundary conditions $\psi\to 0$
toward $r\to \infty$
is the trivial solution
$Q_{1,{\rm out}}= 0$,
$Q_{2,{\rm out}}=0$,
$Q_{3,{\rm out}}=0$,
and 
$Q_{4,{\rm out}}=0$.
In the rest,
to obtain the localized solutions,
we exclude the case of $\omega\neq 0$.

For $\omega=0$,
the exact solution can be obtained as 
\begin{eqnarray}
Q_{1,{\rm out}}
&=&
{\rm exp}
\left[
-2
{\rm arccosh}
\left(
\frac{c\sqrt{r}}{\sqrt{2GM_0 }}
\right)
\right]
\left(
-C_{1,{\rm out}}
{\rm exp}
\left[
4
{\rm arccosh}
\left(
\frac{c\sqrt{r}}{\sqrt{2GM_0 }}
\right)
\right]
-C_{2,{\rm out}}
\right),
\nonumber\\
Q_{2,{\rm out}}
&=&
{\rm exp}
\left[
-2
{\rm arccosh}
\left(
\frac{c\sqrt{r}}{\sqrt{2GM_0 }}
\right)
\right]
\left(
C_{1,{\rm out}}
{\rm exp}
\left[
4
{\rm arccosh}
\left(
\frac{c\sqrt{r}}{\sqrt{2GM_0 }}
\right)
\right]
-C_{2,{\rm out}}
\right),
\nonumber
\\
Q_{3,{\rm out}}
&=&
{\rm exp}
\left[
-2
{\rm arccosh}
\left(
\frac{c\sqrt{r}}{\sqrt{2GM_0 }}
\right)
\right]
\left(
-C_{3,{\rm out}}
{\rm exp}
\left[
4
{\rm arccosh}
\left(
\frac{c\sqrt{r}}{\sqrt{2GM_0 }}
\right)
\right]
-C_{4,{\rm out}}
\right),
\nonumber
\\ 
Q_{4,{\rm out}}
&=&
{\rm exp}
\left[
-2
{\rm arccosh}
\left(
\frac{c\sqrt{r}}{\sqrt{2GM_0 }}
\right)
\right]
\left(
C_{3,{\rm out}}
{\rm exp}
\left[
4
{\rm arccosh}
\left(
\frac{c\sqrt{r}}{\sqrt{2GM_0 }}
\right)
\right]
-C_{4,{\rm out}}
\right),
\end{eqnarray}
where $C_{1,{\rm out}}$, $C_{2,{\rm out}}$, $C_{3,{\rm out}}$, and $C_{4,{\rm out}}$
are integration constants,
whose asymptotic structure at the spatial infinity
is then given by 
\begin{eqnarray}
Q_{1,{\rm out}}
&=&
-\frac{2c^2 C_{1,{\rm out}}}
         {GM_0}
r
+2C_{1,{\rm out}} 
+
\frac{\left(
  C_{1,{\rm out}}
-C_{2,{\rm out}}
        \right)GM_0}
        {2c^2r}
+{\cal O} \left(\frac{1}{r^2}\right),
\nonumber\\
Q_{2,{\rm out}}
&=&
\frac{2c^2 C_{1,{\rm out}}}
         {GM_0}
r
-2C_{1,{\rm out}} 
-
\frac{\left(
C_{1,{\rm out}}
+C_{2,{\rm out}}
        \right)GM_0}
        {2c^2r}
+{\cal O} \left(\frac{1}{r^2}\right),
\nonumber
\\
Q_{3,{\rm out}}
&=&
-\frac{2c^2 C_{3,{\rm out}}}
         {GM_0}
r
+2C_{3,{\rm out}} 
+
\frac{\left(
  C_{3,{\rm out}}
-C_{4,{\rm out}}
        \right)GM_0}
        {2c^2r}
+{\cal O} \left(\frac{1}{r^2}\right),
\nonumber
\\ 
Q_{4,{\rm out}}
&=&
\frac{2c^2 C_{3,{\rm out}}}
         {GM_0}
r
-2C_{3,{\rm out}} 
-
\frac{\left(
  C_{3,{\rm out}}
+C_{4,{\rm out}}
        \right)GM_0}
        {2c^2r}
+{\cal O} \left(\frac{1}{r^2}\right).
\end{eqnarray}
The boundary conditions $\psi_{1,{\rm out}}\to 0$
and $\psi_{2,{\rm out}}\to 0$ as $r\to \infty$ 
impose $C_{1,{\rm out}} =C_{3,{\rm out}} =0$,
and then 
the solutions of $Q_{k,{\rm out}}$ ($k=1,2,3,4$) reduce to 
\begin{eqnarray}
\label{external_dirac_p}
\psi_{1,{\rm out}}
&=&
-
\frac{
\left(
C_{2,{\rm out}} +i C_{4,{\rm out}}
\right)
e^{i \frac{\phi}{2}}
{\rm exp}
\left[
-2 {\rm arccosh}
\left(
\frac{c\sqrt{r}}{\sqrt{2GM_0}}
\right)
\right]
}
        {r e^{\nu/4}}
\begin{pmatrix}
\cos\left(\frac{\theta}{2}\right)
(1+i)
\\ 
\sin\left(\frac{\theta}{2}\right)
(1+i) 
\\
\cos\left(\frac{\theta}{2}\right)
(1-i)
\\
\sin\left(\frac{\theta}{2}\right)
(1-i) 
\end{pmatrix},
\\
\label{external_dirac_m}
\psi_{2,{\rm out}}
&=&
-
\frac{\left(
C_{2,{\rm out}} +i C_{4,{\rm out}}
\right)
e^{-i \frac{\phi}{2}}
{\rm exp}
\left[
-2 {\rm arccosh}
\left(
\frac{c\sqrt{r}}{\sqrt{2GM_0}}
\right)
\right]}
        {r e^{\nu/4}}
\begin{pmatrix}
-
\sin\left(\frac{\theta}{2}\right)
(1-i)
\\ 
\cos\left(\frac{\theta}{2}\right)
(1-i) 
\\
\sin\left(\frac{\theta}{2}\right)
(1+i)
\\
-
 \cos\left(\frac{\theta}{2}\right)
(1+i) 
\end{pmatrix}.
\end{eqnarray}
The interior and exterior solutions,
Eqs. 
\eqref{internal_dirac_p} and \eqref{internal_dirac_m} (with $\omega=0$)
and 
Eqs. 
\eqref{external_dirac_p} and \eqref{external_dirac_m},
respectively,
are matched at the surface of the star $r=\mathcal{R}$,
which gives the relations
\begin{eqnarray}
\label{matching}
{C}_{2,{\rm out}}
&=&
-
Q_{1,{\rm in}}(\mathcal{R})
{\rm exp}
\left(
2{\rm arccosh}
\left(\frac{c\sqrt{\mathcal{R}}}
               {\sqrt{2G M_0}}
\right)
\right)
=
-
Q_{2,{\rm in}}(\mathcal{R})
{\rm exp}
\left(
2{\rm arccosh}
\left(\frac{c\sqrt{\mathcal{R}}}
               {\sqrt{2G M_0}}
\right)
\right),
\nonumber
\\
{C}_{4,{\rm out}}
&=&
- 
Q_{3,{\rm in}}(\mathcal{R})
{\rm exp}
\left(
2{\rm arccosh}
\left(\frac{c\sqrt{\mathcal{R}}}
               {\sqrt{2G M_0}}
\right)
\right)
=
-
Q_{4,{\rm in}}(\mathcal{R})
{\rm exp}
\left(
2{\rm arccosh}
\left(\frac{c\sqrt{\mathcal{R}}}
               {\sqrt{2G M_0}}
\right)
\right),
\end{eqnarray}
which lead to 
the conditions at the surface of the star $r=\mathcal{R}$
$Q_{1,{\rm in}}(\mathcal{R}) =Q_{2,{\rm in}}(\mathcal{R})$,
and 
$Q_{3,{\rm in}}(\mathcal{R})  =Q_{4,{\rm in}}(\mathcal{R})$.
The problem is now to find the coupling constant $\beta_1$ 
that satisfies these conditions,
and 
then $C_{2,{\rm out}}$ and $C_{4,{\rm out}}$
can be evaluated via Eq.  \eqref{matching}.
The full solutions $Q_k (r)$ ($k=1,2,3,4$) in the entire spacetime
can be constructed as
\begin{eqnarray}
Q_k(r) = \left\{
\begin{array}{ll}
Q_{k,{\rm in}}   (r)  & (0<r <\mathcal{R})\\
Q_{k,{\rm out}}  (r)& (r>\mathcal{R})
\end{array}
\right.,
\end{eqnarray}
with the above matching conditions.

\subsection{Numerical solutions}

\begin{figure}[h]
\unitlength=1.1mm
\begin{center}
  \includegraphics[height=4.5cm,angle=0]{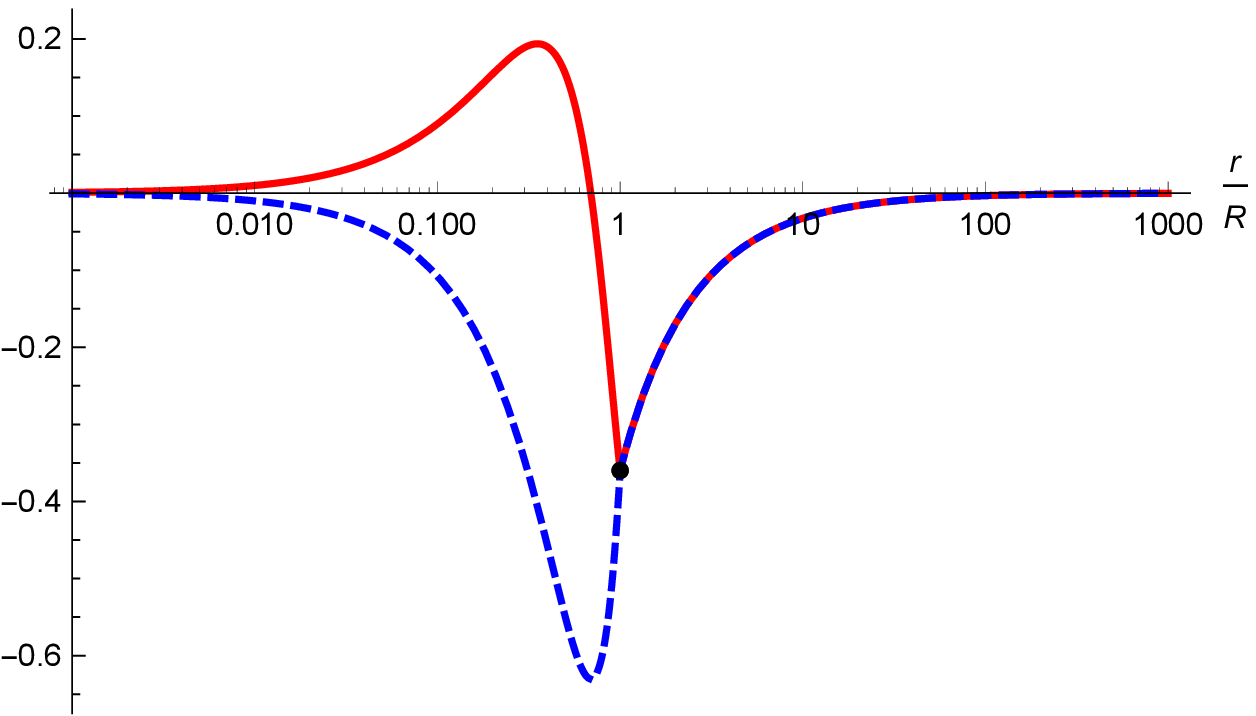}  
  \includegraphics[height=4.5cm,angle=0]{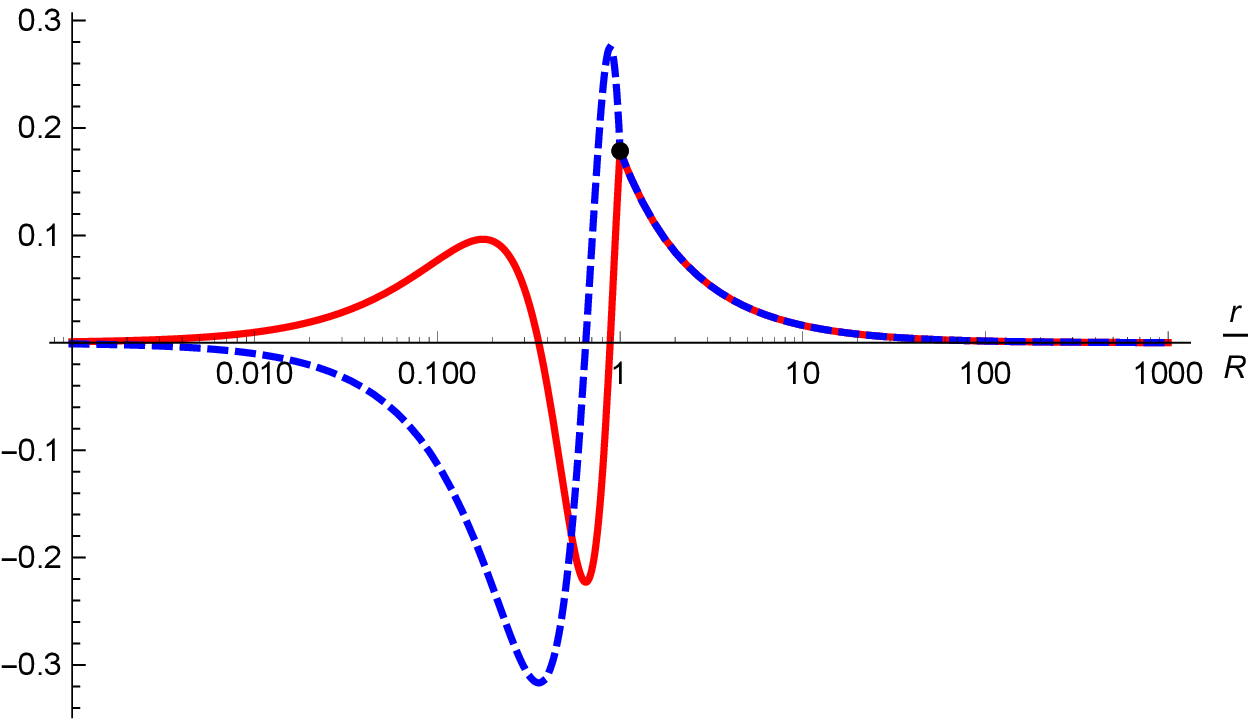}
  \includegraphics[height=4.5cm,angle=0]{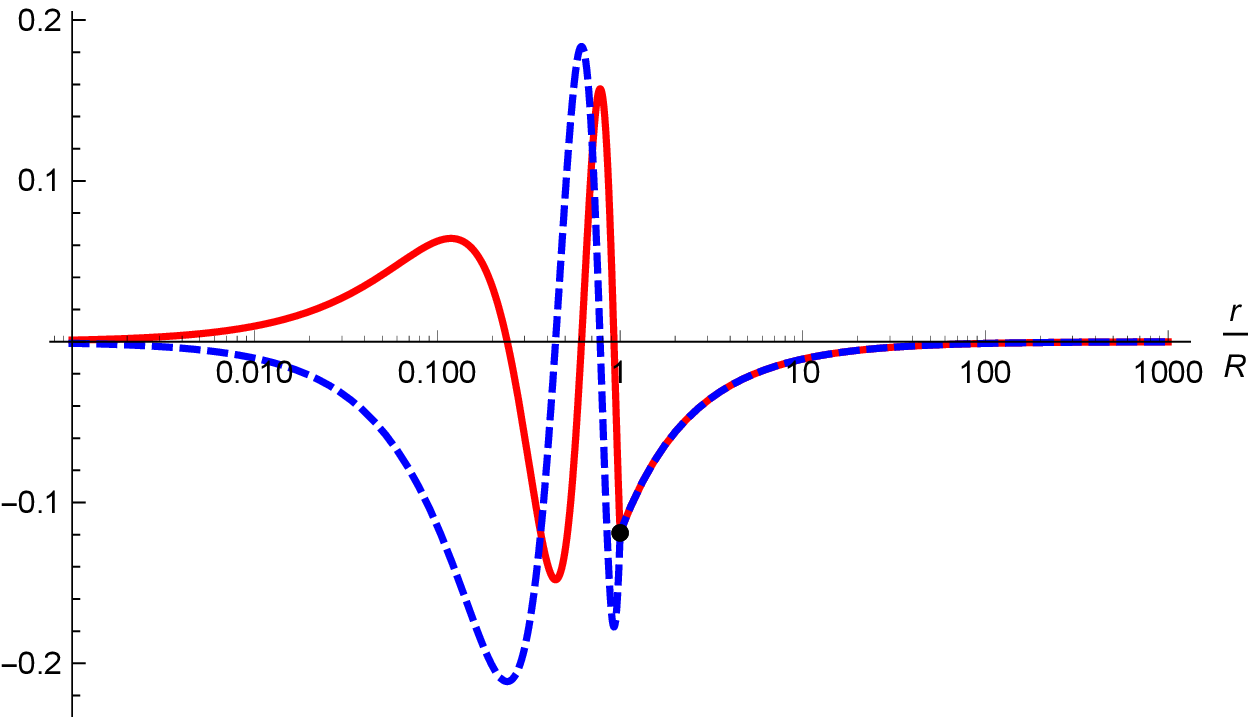}
  \includegraphics[height=4.5cm,angle=0]{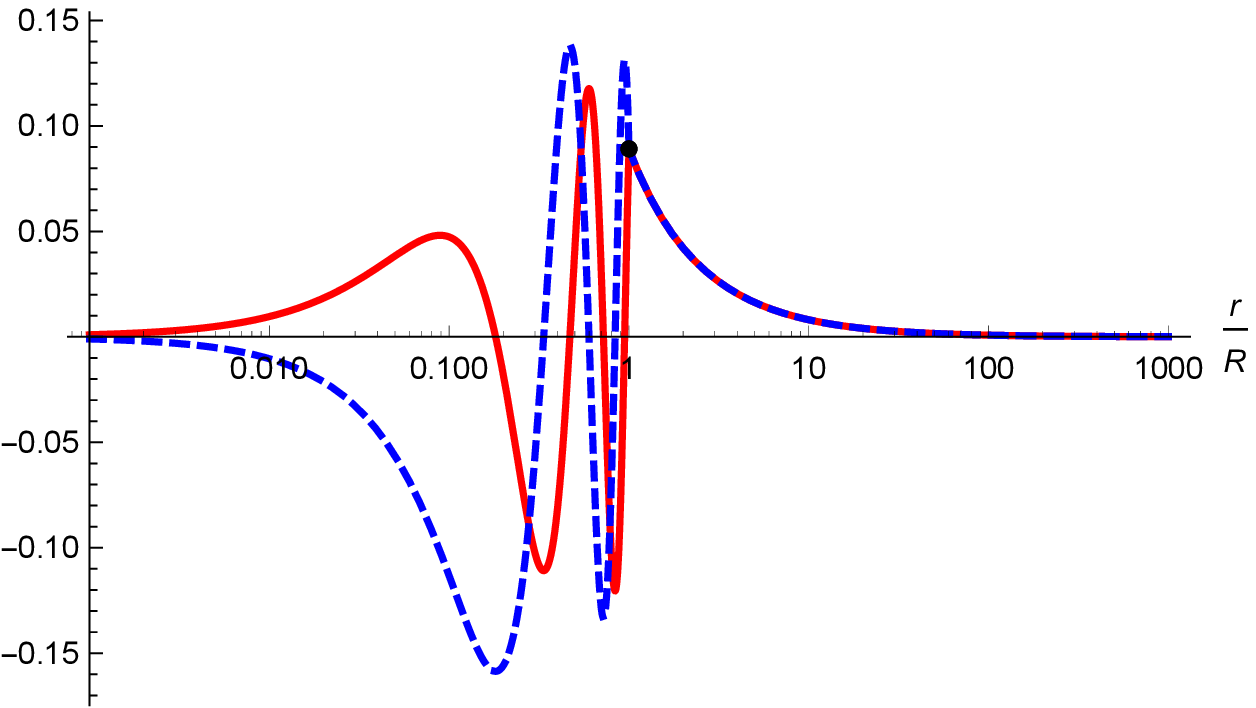}
\caption{
$Q_1$ and $Q_2$ are shown as the functions of $x$ for ${\cal C}=0.1$.
The top-left, top-right, bottom-left, and bottom-right panels correspond 
to the 0-node, 1-node,  2-node, and 3-node solutions of $Q_2$
obtained for $\bar{\beta}_1=11.32, 22.95, 34.50, 46.04$,
respectively.
The red and blue dashed curves
correspond to $Q_1$ and $Q_2$,
respectively.
The black points correspond to the surface of the star at $x=1$.
$Q_1$ always has one more node than $Q_2$.
}
  \label{spinorm}
\end{center}
\end{figure}

We then investigate numerical solutions
satisfying the conditions
$Q_{1,{\rm in}}(\mathcal{R}) =Q_{2,{\rm in}}(\mathcal{R})$
and 
$Q_{3,{\rm in}}(\mathcal{R})  =Q_{4,{\rm in}}(\mathcal{R})$.
For this purpose, 
we introduce the dimensionless quantities
$\bar{\beta}_1
:=c^4 \beta_1/(8\pi  G {\cal R})$
and $x:=r/\mathcal{R}$,
where $x=1$ corresponds to the surface of the star.
In Fig. \ref{spinorm},
$Q_1$ and $Q_{2}$ are shown as the functions of $x$ for ${\cal C}=0.1$.
The top-left, top-right, bottom-left, and bottom-right panels correspond 
to the 0-, 1-, 2-,  and 3-node solutions of $Q_2$
which are 
obtained for $\bar{\beta}_1=11.32, 22.95, 34.50, 46.04$,
respectively.
The red and blue dashed curves
correspond to $Q_1$ and $Q_2$,
respectively.
The black points correspond to the surface of the star at $x=1$.
As $\bar{\beta}_1$ increases,
the solutions with more nodes are obtained.
We note that 
$Q_1$ always has one more node than $Q_2$.
As expected,
the first order derivatives $Q_1'(x)$ and $Q_2'(x)$ 
are discontinuous at the surface of the star $x=1$ 
for any number of nodes.
We note that 
since the equations are linear,
if $Q_1$ and $Q_2$ are solutions, 
$c' Q_1$ and $c' Q_2$ with $c'$ being constant
are also the solutions,
and  
the solutions for $Q_3$ and $Q_4$
can be obtained 
by the relations $Q_3/Q_1=Q_4/Q_2$,
respectively.

We expect that
the above lowest $0$-node solution arises 
as the consequence of the tachyonic instability of 
the trivial solution $\psi=0$.
When we consider the small perturbations 
about the trivial solution $\psi=0$,
the zero frequency eigenstate of the perturbations cannot be the ground state
for ${\bar \beta}_1\geq 11.32$
and the end point of the tachyonic instability 
would be the nontrivial solutions
constructed in this section.

\subsection{Coupling to the pressure}

\begin{figure}[h]
\unitlength=1.1mm
\begin{center}
  \includegraphics[height=4.5cm,angle=0]{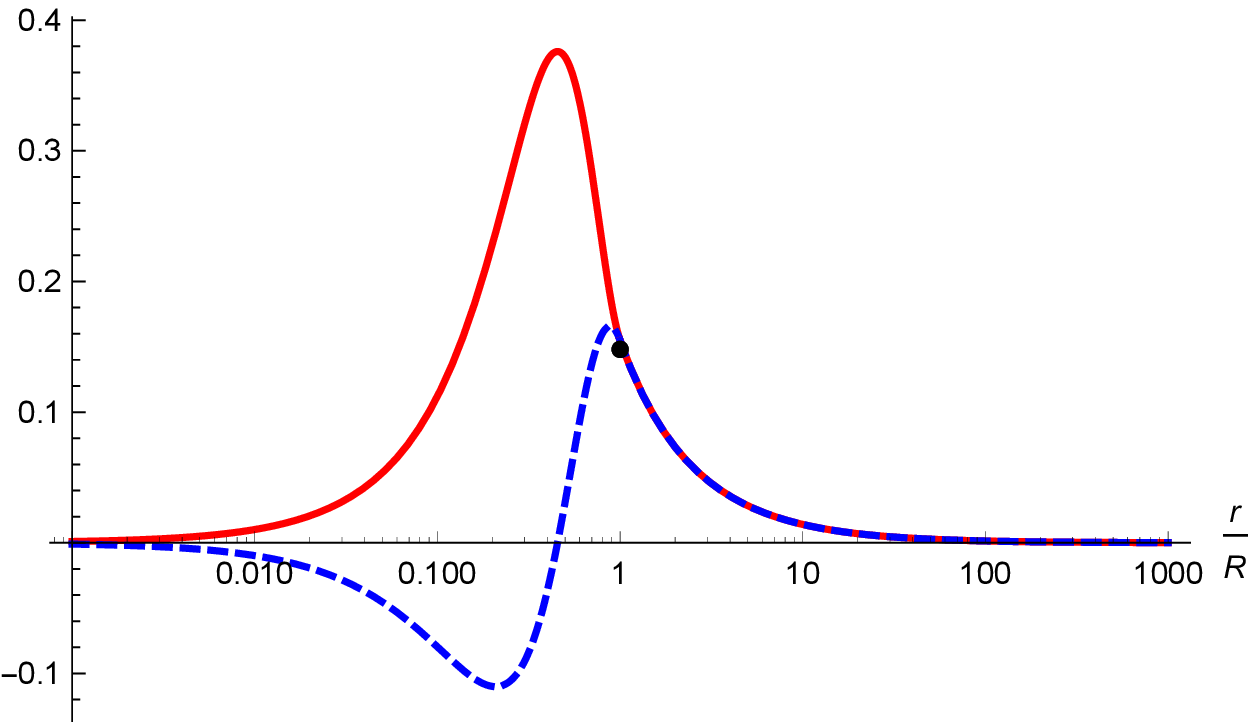}  
  \includegraphics[height=4.5cm,angle=0]{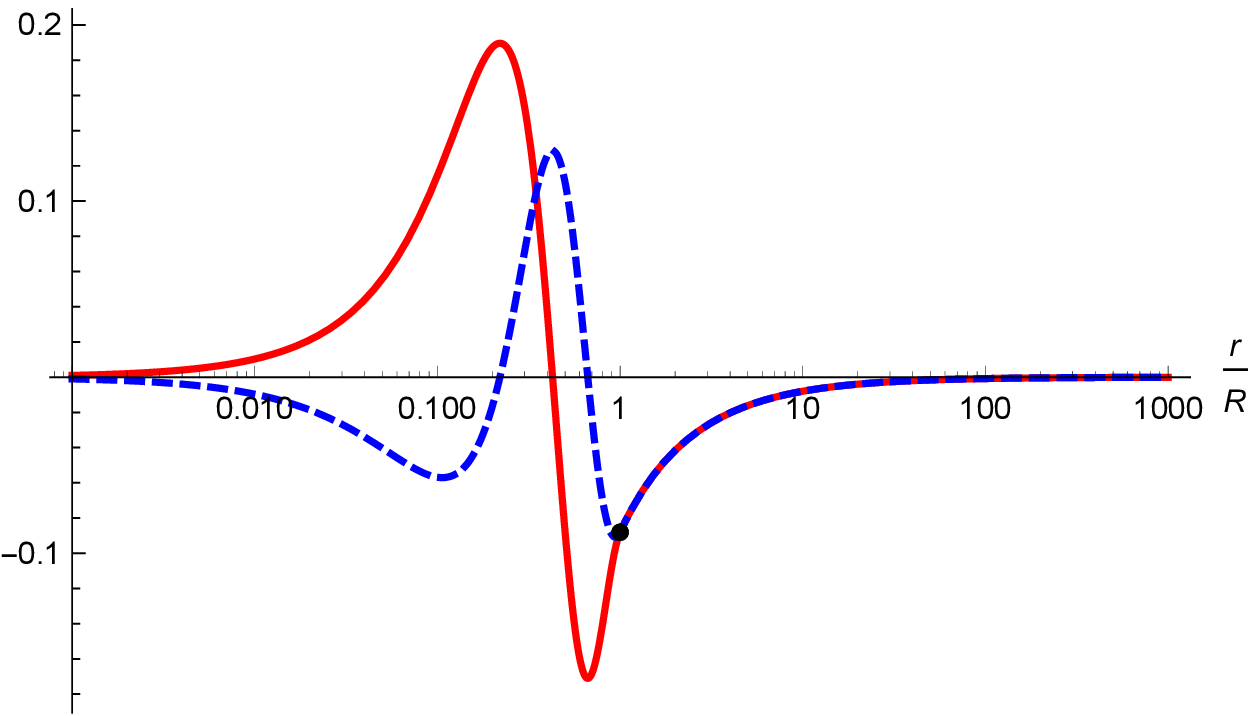}
  \includegraphics[height=4.5cm,angle=0]{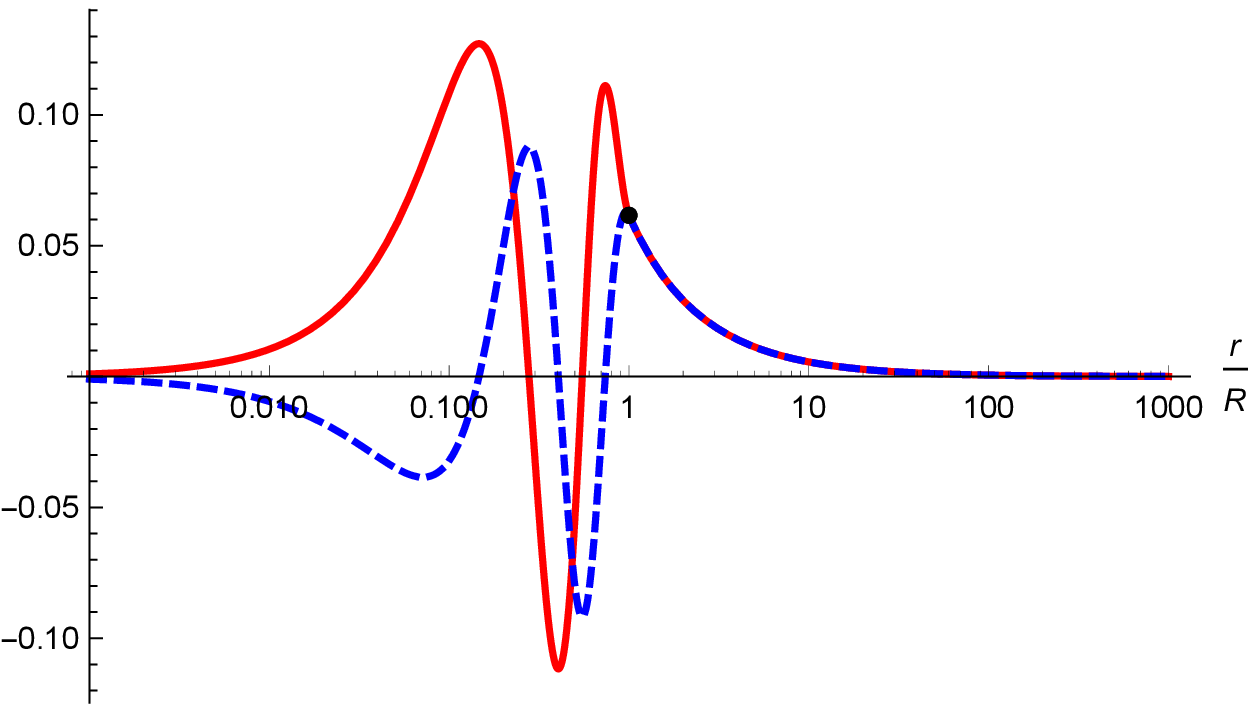}
  \includegraphics[height=4.5cm,angle=0]{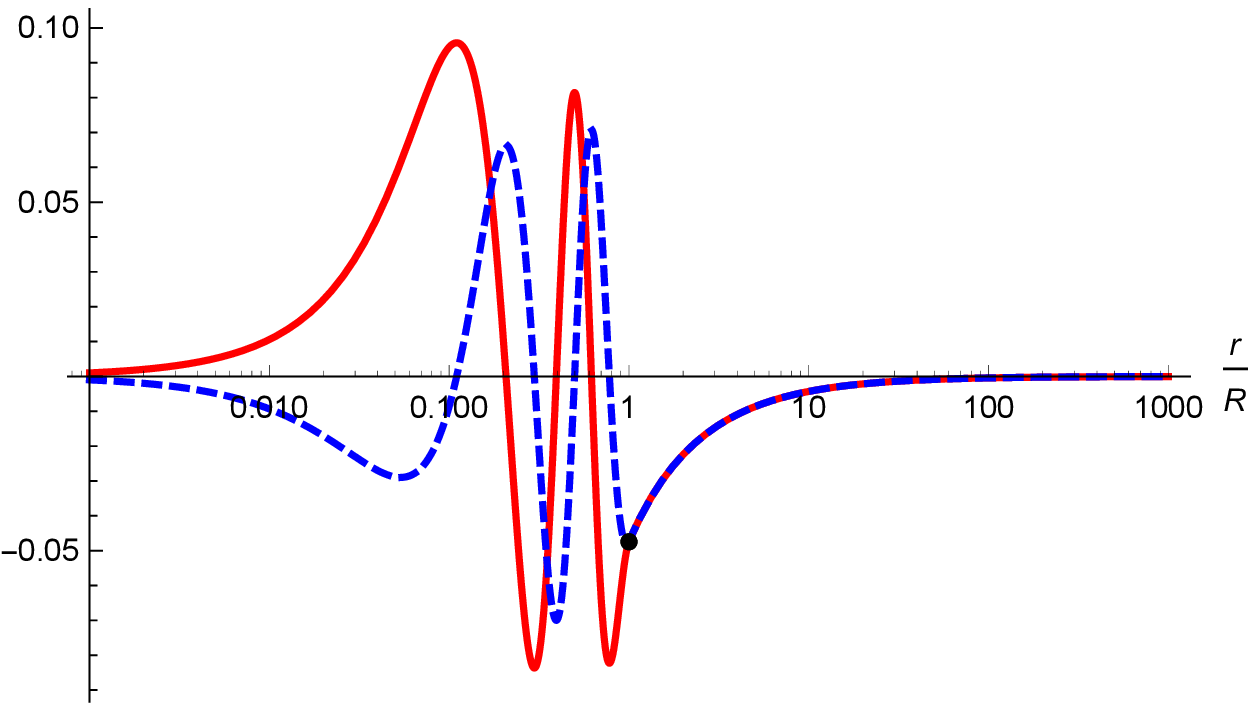}
\caption{
$Q_{1}$ and $Q_{2}$ are shown as the functions 
of $x=r/\mathcal{R}$
for ${\cal C}=0.1$.
The top-left, top-right, bottom-left, and bottom-right panels correspond 
to the 0-, 1-, 2-, and 3-node solutions of $Q_1$
obtained for $\bar{\beta}_2=89.05, 168.82,  249.66, 330.88$,
respectively.
The red and blue dashed curves
correspond to $Q_{1}$ and $Q_{2}$,
respectively.
The black points correspond to the surface of the star.
$Q_2$ always has one more node than $Q_1$.
}
  \label{spinormp}
\end{center}
\end{figure}

The problem of the discontinuity of the first order derivatives
may be solved,
for instance, 
if we consider the coupling only to the pressure
 \begin{eqnarray}
\label{historia2}
\left(
{\gamma}^\mu 
D_\mu 
+
\frac{3\beta_2}{2}
{\hat \gamma}^5
{\tilde p}(r)
\right)
\psi
=0,
\end{eqnarray}
where $\beta_2$ is the coupling constant.
Since the pressure ${\tilde p}(r)$ vanishes at the surface of the star 
$r=\mathcal{R}$,
the first order derivative of $\psi$ with respect to $r$
as well as $\psi$ itself
is continuous there.
Introducing the dimensionless quantity
$\bar{\beta}_2
:=c^4 \beta_2/(8\pi  G {\cal R})$,
in Fig. \ref{spinormp},
we show an example of  $Q_1$ and $Q_{2}$ 
as the functions of $x=r/\mathcal{R}$
for ${\cal C}=0.1$.
The top-left, top-right, bottom-left, and bottom-right panels correspond 
to the 0-, 1-, 2-, and 3-node solutions of $Q_1$
obtained for $\bar{\beta}_2=89.05, 168.82, 249.66, 330.88$,
respectively.
The red and blue dashed curves
correspond to $Q_{1}$ and $Q_{2}$,
respectively.
The black points correspond to the surface of the star.
$Q_2$ always has one more node than $Q_1$.
We note that 
as in the previous case
the solution for $Q_3$ and $Q_4$
can be obtained 
with the relation $Q_3/Q_1=Q_4/Q_2$, 
respectively.

We confirmed that 
the solutions for the Dirac spinor field and 
their first order derivatives 
for any number of nodes
are continuous across the surface 
of the star $x=1$, i.e., $r=\mathcal{R}$.
Thus,
any component of the effective energy-momentum tensor 
is continuous across the surface of the star,
even if it takes a nonzero value.
However, 
since Eq. \eqref{historia2}
is not derived from the variation of the action
and is more specific to the observer who is at rest at the spatial infinity,
the above argument seems to depend on the choice of the coordinates.
Further studies on the possible improved couplings 
would be left for future work.

\section{Stealth spontaneous spinorization}
\label{sec5}

One might imagine that 
the discontinuity of the first order derivative of
$Q_1$, $Q_2$, $Q_3$, and $Q_4$
at the surface of the star $x=1$,
$r=\mathcal{R}$
would  result in
the discontinuity 
in the effective energy density and pressure 
of the Dirac spinor field.
If it is the case,
there should be the localized source 
of the Dirac spinor field at the surface of the star,
whose existence seems to be unphysical.
We again emphasize that 
the discontinuity is somewhat an artifact
of the idealization of the background solutions,
and
since 
in realistic relativistic stellar backgrounds
the matter energy density would quickly but smoothly go to zero,
the discontinuity of $\partial_r \psi$ would also be smoothened.
In addition,
in this section, 
we argue
that even in the original model
the existence of  the discontinuity does not matter,
as all the components of the effective energy-momentum tensor 
of the Dirac field 
trivially vanish in any spherically symmetric backgrounds.

We now define 
the effective energy-momentum tensor for the Dirac field 
from Eqs. \eqref{eq_model_a} and \eqref{coupling},
which up to the quadratic order of $\psi$ and ${\bar \psi}$
is given by 
\begin{eqnarray}
T^{(\psi,{\rm eff})}_{\mu\nu}
&:=&
-
\frac{1}{2}
\sum_{I=1,2}
\left[
{\bar \psi}_I
{\hat \gamma}^5
e_{b(\mu}
{\hat\gamma}^b
D_{\nu)}
\psi_I
-
\left(
D_{(\mu}
{\bar \psi}_I
\right)
{\hat \gamma}^5
{\hat\gamma}^b
e_{b |\nu)}
\psi_I
\right]
+
\beta_1
{T}^{(m)}_{\mu\nu}
\sum_{I=1,2}
{\bar\psi}_I
\psi_I
+{\cal O} (\psi^4),
\end{eqnarray}
where ${\cal O} (\psi^4)$ denotes all the quartic order combinations
of $\psi_I$, ${\bar\psi}_I$ and their first order derivatives.
Following the arguments in Sec. \ref{sec32},
it is straightforward 
to confirm 
that for any two Dirac spinor fields
that maintain the spherical symmetry of the spacetime
\begin{eqnarray}
\label{ansatz_dirac}
\psi_1
=
\frac{h(t) e^{i \frac{\phi}{2}}}
        {r e^{\nu/4}}
\begin{pmatrix}
\cos\left(\frac{\theta}{2}\right)
\left[
(1+i) f(r)+ (1-i)g(r)
\right]
\\ 
i\sin\left(\frac{\theta}{2}\right)
\left[
(1-i) f(r)+ (1+i)g(r)
\right]
\\
\cos\left(\frac{\theta}{2}\right)
\left[
(1-i) f(r)+ (1+i)g(r)
\right]
\\ 
-i \sin\left(\frac{\theta}{2}\right)
\left[
(1+i) f(r)+ (1-i)g(r)
\right]
\end{pmatrix},
\end{eqnarray}
and 
\begin{eqnarray}
\label{ansatz_dirac2}
\psi_2
=
\frac{h(t)e^{-i \frac{\phi}{2}}}
        {r e^{\nu/4}}
\begin{pmatrix}
i
\sin\left(\frac{\theta}{2}\right)
\left[
(1+i) f(r)+ (1-i)g(r)
\right]
\\ 
\cos\left(\frac{\theta}{2}\right)
\left[
(1-i) f(r)+ (1+i)g(r)
\right]
\\
i
\sin \left(\frac{\theta}{2}\right)
\left[
(1-i) f(r)+ (1+i)g(r)
\right]
\\ 
-\cos\left(\frac{\theta}{2}\right)
\left[
(1+i) f(r)+ (1-i)g(r)
\right]
\end{pmatrix},
\end{eqnarray}
where $h(t)$ is any complex function of time $t$, 
all the components of $T^{(\psi,{\rm eff})}_{\mu\nu}$
vanish
both inside and outside the star
at least at ${\cal O}(\psi^2)$.
The limits to the surface of the star from the inside  ($r\to {\cal R}_{-0}$)
and that from the outside ($r\to {\cal R}_{+0}$)
converge to the unique value $0$.
Thus, all components of the effective energy-momentum tensor 
for the Dirac spinor  vanish in the entire spacetime
at least at ${\cal O}(\psi^2)$.
This indicates 
that the tachyonic growth of the Dirac spinor field does 
not backreact on the spacetime geometry,
and the metric solution remains the same as in general relativity.
The tachyonic growth 
would be quenched
by the nonlinear effects in 
the effective Dirac field equation \eqref{Dirac_inside}
as in the case of spontaneous scalarization~\cite{Minamitsuji:2018xde,Silva:2018qhn}.
As argued in Sec. \ref{sec33},
since $\rho_{(\psi_2)} =-\rho_{(\psi_1)}$,
one of the two Dirac fields has the positive energy density,
while the other has the negative one.
However,  the negative energy density of one field does not  imply any pathology,
as only the combination of the two Dirac fields has the
physical meaning to maintain the static and spherically symmetric 
spacetime. 
Since $\Phi=\sum_I {\bar \psi}_I \psi_I =0$, $F=1$ in Eq. \eqref{coupling}
and hence there is no distinction between the Jordan and Einstein frames.
After all, we expect that 
spontaneous spinorization
proceeds as a stealth process 
and 
does not leave any observable effects.
Thus, we call this stealth spinorization.

Since 
the property of 
the vanishing effective energy-momentum tensor 
holds 
in any static and spherically symmetric spacetime,
we conclude that
the above stealth spinorization
will happen 
in any static and spherically symmetric general relativistic stellar backgrounds
with more realistic equations of state.
As the next step,  
it is of interest 
whether the conclusion remains the same 
in less symmetric general relativistic stellar backgrounds,
such as 
stationary and axisymmetric spacetimes around rotating stars.
Moreover,
since our model introduced in Sec. \ref{sec2}
corresponds to the simplest model for spontaneous spinorization,
it is also of interest 
to explore more general models \cite{Ramazanoglu:2018hwk}.

\section{Conclusions}
\label{sec6}

We have investigated the possibility of the tachyonic growth of the Dirac spinor field $\psi$
on general relativistic stellar backgrounds and
spontaneous spinorization of compact objects,
which is analogous to spontaneous scalarization
studied in many previous works. 
We have focused on the theory \eqref{einstein_dirac_action} and \eqref{kin},
which is composed of 
the modified kinetic term by the insertion of the fifth gamma matrix ${\hat \gamma}^5$
and the conformal coupling of the Dirac spinor field to matter
and would
lead to the tachyonic growth of the Dirac spinor field only in the high density backgrounds.
At the linearized level,
the Dirac equation \eqref{Dirac_tachyonic}
is very similar to the tachyonic Dirac spinor field theory \eqref{tachyonic_flat},
where the mass parameter $M$
is replaced by the trace of the matter energy-momentum tensor $T^{(m)\mu}{}_\mu$.

As in the case of  spontaneous scalarization,
we naively expect
that the tachyonic growth of the Dirac spinor field 
would significantly
modify the structure of relativistic stars from general relativistic ones.
In order to obtain the spherically symmetric solutions,
we have to consider the two Dirac fields at the same time
such as Eqs.~\eqref{internal_dirac_p} and \eqref{internal_dirac_m}.
We have confirmed that
at the linearized equation for $\psi$
our theory gives rise to the nontrivial profiles of the Dirac spinor field
with any number of nodes and the vanishing field values at the spatial infinity. 
However,
we have also confirmed that 
all the components of the effective energy-momentum tensor 
of the Dirac spinor fields
vanish 
up to the quadratic order of $\psi$ and ${\bar \psi}$
in the entire spacetime including the surface of the star,
when the contributions of both the two Dirac spinor fields 
are taken into consideration.
Thus,
we have expected 
that spontaneous spinorization 
proceeds as a stealth process unlike spontaneous scalarization,
and hence does not leave any observable effects.

The status of the study of spontaneous spinorization
is still immature,
and 
several issues should be clarified in future work.
Although we have focused on the static and spherically symmetric stars,
it would be important to clarify
whether this stealth property holds
in less symmetric backgrounds
such as rotating and binary stars.
It would also be necessary to 
follow the time evolution from 
general relativistic stellar solutions 
to those with nontrivial profile of the Dirac spinor fields.
Another direction of the study 
is to generalize the model considered in this paper
and explore novel features of spontaneous spinorization.
We hope to come back to these issues in our future work.

\acknowledgments{
M.M.~was supported by the Portuguese national fund 
through the Funda\c{c}\~{a}o para a Ci\^encia e a Tecnologia
within the framework of the Decree-Law 57/2016  of August 29
(changed by Law 57/2017 of July 19),
and the CENTRA through the Project~No.~UIDB/00099/2020.
}

\appendix


\section{
Consistent action for the Dirac spinor field in the flat and curved spacetimes}
\label{app_a}

\subsection{In the Minkowski spacetime}

In this appendix, we review the consistency of the equation of motion
for the Dirac adjoint $\bar{\psi}$ in the tachyonic Dirac field theory.
Varying the actions 
\eqref{tachyonic_flat} 
and 
\eqref{tachyonic_flat_alt}
with respect to $\bar\psi$ and $\psi$,
we obtain the following equations,
respectively:
\begin{eqnarray}
\label{eq1}
\left(
{\hat\gamma}^5{\hat\gamma}^a
\partial_a
-M
\right)
\psi=0,
\qquad 
{\bar\psi}
\left(
\overleftarrow{\partial}_a
{\hat \gamma}^5
{\hat \gamma}^a
+M
\right)
=0,
\end{eqnarray}
and
\begin{eqnarray}
\label{eq2}
\left(
{\hat\gamma}^5{\hat\gamma}^a
\partial_a
-M
\right)\psi=0,
\qquad 
{\bar\psi}
\left(
\overleftarrow{\partial}_a
{\hat \gamma}^a
{\hat \gamma}^5
+M
\right)
=0.
\end{eqnarray}
Since $\{{\hat \gamma}^5,{\hat \gamma}^a\}=0$,
we obtain the different results in the equations for $\bar \psi$,
and see which one is consistent.
Taking the conjugate transpose of the equation for $\psi$,
\begin{eqnarray}
0
&=&
{\psi}^\dagger
\left(
\overleftarrow{\partial}_a 
\left({\hat \gamma}^a\right)^\dagger
{\hat \gamma}^5
-M
\right)
=
{\psi}^\dagger
\left(
-
\overleftarrow{\partial}_a 
{\hat \gamma}^5
\left({\hat \gamma}^a\right)^\dagger
-M
\right)
\nonumber\\
&=&
{\psi}^\dagger
\left(
-
\overleftarrow{\partial}_0
{\hat \gamma}^5
\left({\hat \gamma}^0\right)^\dagger
-
\overleftarrow{\partial}_i
{\hat \gamma}^5
\left({\hat \gamma}^i\right)^\dagger
-M
\right)
=
{\psi}^\dagger
\left(
-
\overleftarrow{\partial}_0
{\hat \gamma}^0
{\hat \gamma}^5
+
\overleftarrow{\partial}_i
{\hat \gamma}^i
{\hat \gamma}^5
-M
\right),
\end{eqnarray}
where the index $i=1,2,3$ represent the spatial directions
and we have employed
$({\hat\gamma}^5)^\dagger={\hat\gamma}^5$,
$
{\hat \gamma}^5
({\hat \gamma}^0)^\dagger
-
{\hat \gamma}^0
{\hat \gamma}^5
=0$,
and 
$
{\hat \gamma}^5
({\hat \gamma}^i)^\dagger
+
{\hat \gamma}^i
{\hat \gamma}^5
=0$.
Multiplying $-i{\hat \gamma}^0$ from the right side,
\begin{eqnarray}
0
&=&
-i
{\psi}^\dagger
\left(
-
\overleftarrow{\partial}_0
{\hat \gamma}^0
{\hat \gamma}^5
+
\overleftarrow{\partial}_i
{\hat \gamma}^i
{\hat \gamma}^5
-M
\right)
{\hat \gamma}^0
=
-i
{\psi}^\dagger
\left(
-
\overleftarrow{\partial}_0
{\hat \gamma}^0
{\hat \gamma}^5
+
\overleftarrow{\partial}_i
{\hat \gamma}^i
{\hat \gamma}^5
\right)
{\hat \gamma}^0
-M
{\bar \psi}
\nonumber\\
&=&
-i
{\psi}^\dagger
\left(
\overleftarrow{\partial}_0
{\hat \gamma}^0
{\hat \gamma}^0
{\hat \gamma}^5
+
\overleftarrow{\partial}_i
{\hat \gamma}^0
{\hat \gamma}^i
{\hat \gamma}^5
\right)
-M
{\bar \psi}
=
{\bar\psi}
\left(
\overleftarrow{\partial}_0
{\hat \gamma}^0
{\hat \gamma}^5
+
\overleftarrow{\partial}_i
{\hat \gamma}^i
{\hat \gamma}^5
\right)
-M
{\bar \psi}
\nonumber\\
&=&
{\bar\psi}
\left(
\overleftarrow{\partial}_a
{\hat \gamma}^a
{\hat \gamma}^5
-M
\right)
=
-
{\bar\psi}
\left(
\overleftarrow{\partial}_a
{\hat \gamma}^5
{\hat \gamma}^a
+M
\right),
\end{eqnarray}
which reproduces the equation for ${\bar \psi}$ in Eq. \eqref{eq1}.
Thus, the action \eqref{tachyonic_flat}
provides the consistent equation of motion for $\bar\psi$
\cite{Jentschura:2011ga,Ramazanoglu:2018hwk}.

\subsection{In the curved spacetimes}

The extension to the case of the curved spacetime is also straightforward.
Varying the action of  the tachyonic Dirac field \eqref{lag_tachyon}
in the curved spacetime 
with respect to $\bar\psi$ and $\psi$,
respectively
we obtain the following equations:
\begin{eqnarray}
\label{app^5}
\left(
{\hat\gamma}^5
{\gamma}^\mu
\left(
\partial_\mu
+\Gamma_\mu
\right)
-M
\right)\psi=0,
\qquad 
{\bar\psi}
\left(
\left(
\overleftarrow{\partial}_\mu
-\Gamma_\mu
\right)
{\hat \gamma}^5
{\gamma}^\mu
+M
\right)
=0.
\end{eqnarray}
Taking the conjugate transpose to the equation for $\psi$,
\begin{eqnarray}
0
&=&
{\psi}^\dagger
\left(
\left(
 \overleftarrow{\partial}_\mu
+(\Gamma_\mu)^\dagger
\right) 
\left({\gamma}^\mu\right)^\dagger
{\hat \gamma}^5
-M
\right)
=
{\psi}^\dagger
\left(
\left(
 \overleftarrow{\partial}_t
+(\Gamma_t)^\dagger
\right) 
\left({\gamma}^t\right)^\dagger
{\hat \gamma}^5
+
\left(
 \overleftarrow{\partial}_k
+(\Gamma_k)^\dagger
\right) 
\left({\gamma}^k\right)^\dagger
{\hat \gamma}^5
-M
\right)
\nonumber\\
&=&
{\psi}^\dagger
\left(
-
\left(
 \overleftarrow{\partial}_t
+(\Gamma_t)^\dagger
\right) 
{\gamma}^t
{\hat \gamma}^5
+
\left(
 \overleftarrow{\partial}_k
+(\Gamma_k)^\dagger
\right) 
{\gamma}^k
{\hat \gamma}^5
-M
\right),
\end{eqnarray}
where
$k$ denotes the spatial indices
and we have employed
the relations of the gamma matrices
$({\hat\gamma}^5)^\dagger={\hat\gamma}^5$,
$\gamma^t
+
\left( \gamma^t \right)^\dagger
=0$,
and 
$\gamma^k
-\left( \gamma^k \right)^\dagger
=0$.
Multiplying $-i{\hat \gamma}^0$ from the right side,
\begin{eqnarray}
0
&=&
-i
{\psi}^\dagger
\left(
-
\left(
 \overleftarrow{\partial}_t
+(\Gamma_t)^\dagger
\right) 
{\gamma}^t
{\hat \gamma}^5
+
\left(
 \overleftarrow{\partial}_k
+(\Gamma_k)^\dagger
\right) 
{\gamma}^k
{\hat \gamma}^5
-M
\right)
{\hat \gamma}^0
\nonumber\\
&=&
-i
{\psi}^\dagger
\left(
\left(
 \overleftarrow{\partial}_t
+(\Gamma_t)^\dagger
\right) 
{\hat \gamma}^0
{\gamma}^t
{\hat \gamma}^5
+
\left(
 \overleftarrow{\partial}_k
+(\Gamma_k)^\dagger
\right)
{\hat \gamma}^0 
{\gamma}^k
{\hat \gamma}^5
-M{\hat \gamma}^0
\right)
\nonumber\\
&=&
{\bar \psi}
\left(
\left(
 \overleftarrow{\partial}_t
-\Gamma_t
\right) 
{\gamma}^t
{\hat \gamma}^5
+
\left(
 \overleftarrow{\partial}_k
-\Gamma_k
\right)
{\gamma}^k
{\hat \gamma}^5
-M
\right)
=
{\bar \psi}
\left(
\left(
 \overleftarrow{\partial}_\mu
-\Gamma_\mu
\right) 
{\gamma}^\mu
{\hat \gamma}^5
-M
\right)
\nonumber\\
&=&
-
{\bar \psi}
\left(
\left(
 \overleftarrow{\partial}_\mu
-\Gamma_\mu
\right) 
{\hat \gamma}^5
{\gamma}^\mu
+M
\right),
\end{eqnarray}
where we have used 
${\hat\gamma}^0 \Gamma_\mu
+\left( \Gamma_\mu \right)^\dagger
{\hat \gamma}^0
=0$,
which reproduces the equation for ${\bar \psi}$ in Eq. \eqref{app^5}.
Thus, the action \eqref{lag_tachyon}
provides the consistent equation of motion for $\bar\psi$.

\bibliography{disformal_refs}
\end{document}